\documentclass{jfm}

\usepackage{graphicx}
\usepackage{natbib}
\usepackage{amsmath,rotating}
\usepackage{amsfonts}
\usepackage{amssymb}
\usepackage{color}
\usepackage{boolexpr}
\usepackage{soul}
\usepackage[normalem]{ulem}

\usepackage{placeins}

\ifCUPmtlplainloaded \else
  \checkfont{eurm10}
  \iffontfound
    \IfFileExists{upmath.sty}
      {\typeout{^^JFound AMS Euler Roman fonts on the system,
                   using the 'upmath' package.^^J}%
       \usepackage{upmath}}
      {\typeout{^^JFound AMS Euler Roman fonts on the system, but you
                   dont seem to have the}%
       \typeout{'upmath' package installed. JFM.cls can take advantage
                 of these fonts,^^Jif you use 'upmath' package.^^J}%
      }
  \else
  \fi
\fi

\ifCUPmtlplainloaded \else
  \checkfont{msam10}
  \iffontfound
    \IfFileExists{amssymb.sty}
      {\typeout{^^JFound AMS Symbol fonts on the system, using the
                'amssymb' package.^^J}%
       \usepackage{amssymb}%
       \let\le=\leqslant  
       \let\ge=\geqslant  
      }{}
  \fi
\fi

\ifCUPmtlplainloaded \else
  \IfFileExists{amsbsy.sty}
    {\typeout{^^JFound the 'amsbsy' package on the system, using it.^^J}%
     \usepackage{amsbsy}}
    {\providecommand\boldsymbol[1]{\mbox{\boldmath $##1$}}}
\fi

\newcommand\Ma{\mbox{\textit{Ma}}}  
\newcommand\tke{\mbox{\textit{TKE}}}  
\newcommand\imz{\mbox{\textit{IMZ}}}  

\newsavebox{\astrutbox}
\sbox{\astrutbox}{\rule[-5pt]{0pt}{20pt}}

\newcommand\etal{\mbox{\textit{et al.}}}

\newcommand{\grad} {\ensuremath{\nabla\ }}

\newcommand{\bs}{\boldsymbol}

\newcommand{\divg}{\nabla\cdot}

\def\bgk#1{\mbox{\boldmath $#1$}}
\newcommand{\pth}[1]{\left( #1 \right)}				

\ifcase\boolexpr{ 0=1 }
\newcommand\s{\bgroup\markoverwith
{\textcolor{red}{\rule[0.6ex]{2pt}{0.4pt}}}\ULon}
\newcommand{\rev}[1]{\textcolor{red}{#1}}
\else
\newcommand{\s}[1]{}
\newcommand{\rev}[1]{\textcolor{black}{#1}}
\fi

\title[On the Richtmyer-Meshkov instability]{On the Richtmyer-Meshkov instability evolving from a deterministic multimode planar interface}

\author[V. K. Tritschler, B. J. Olson, S. K. Lele, S. Hickel, X. Y. Hu and N. A. Adams]%
{V. K. Tritschler$^{1,2}$%
 \thanks{Email address for correspondence: volker.tritschler@aer.mw.tum.de},\ns
B. J. Olson$^3$
 \thanks{Email address for correspondence: olson45@llnl.gov},\ns
S. K. Lele$^2$, S. Hickel$^1$, X. Y. Hu$^1$ and N. A. Adams$^1$}
\affiliation{$^1$Institute of Aerodynamics and Fluid Mechanics, Technische Universit\"at M\"unchen, 85747 Garching, Germany\\[\affilskip]
$^2$Department of Aeronautics and Astronautics, Stanford University, Stanford, CA 94305, USA\\[\affilskip]
$^3$Lawrence Livermore National Laboratory, Livermore, CA 94550, USA}

\date{?; revised ?; accepted ?. - To be entered by editorial office}
\begin{document}

\maketitle

\begin{abstract}
We investigate the shock-induced turbulent mixing between a light and heavy gas, where a Richtmyer-Meshkov instability (RMI) is initiated by a $\Ma = 1.5$ shock wave. The prescribed  initial conditions define a deterministic multimode interface perturbation between \s{light and heavy gas} \rev{the gases}, which can be imposed exactly for different simulation codes and resolutions to allow for quantitative comparison. Well-resolved large-eddy simulations are performed using two different and \s{independent} \rev{independently developed} numerical methods with the objective of assessing turbulence structures, prediction uncertainties and convergence behaviour. The two numerical methods differ fundamentally with respect to the employed subgrid-scale regularisation,  each representing state-of-the-art approaches to RMI. Unlike previous studies the focus of the present investigation is to quantify uncertainties introduced by the numerical method, as there is strong evidence that subgrid-scale regularisation and truncation errors may have a significant effect on the linear and non-linear stages of the RMI evolution.
Fourier diagnostics reveal that the larger energy containing scales converge rapidly with increasing mesh resolution and thus are in excellent agreement for the two numerical methods. Spectra of gradient \rev{dependent} quantities, such as enstrophy and scalar dissipation rate, show stronger dependencies on the small-scale flow field structures in consequence of truncation error effects, which for one numerical method are dominantly dissipative and for the other dominantly dispersive. 
Additionally, the study reveals details of various stages of RMI, as the flow transitions from large scale, non-linear entrainment, to fully developed turbulent mixing. The growth rates of the mixing zone widths as obtained by the two numerical methods are $\sim t^{7/12}$ before re-shock, and $\sim (t-t_0)^{2/7}$ long after re-shock. The decay rate of turbulence kinetic energy is consistently $\sim (t-t_0)^{-10/7}$ at late times, where the molecular mixing fraction approaches an asymptotic limit $\Theta \approx 0.85$. \rev{The anisotropy measure $\left< a \right>_{xyz}$ approaches an asymptotic limit of $\approx 0.04$ implying that no full recovery of isotropy within the mixing zone is obtained, even after re-shock. }
Spectra of density, turbulence kinetic energy, scalar dissipation rate and enstrophy are presented and show excellent agreement for the resolved scales. Probability density function of the heavy-gas mass fraction and vorticity reveal that the light-heavy gas composition within the mixing zone is accurately predicted, whereas it is more difficult to capture the long-term behaviour of the vorticity.
\s{The established agreement for the large scales of the solution for two fundamentally different numerical methods is unprecedented and serves to establish a reference data set for further numerical analysis of the RMI evolution.} 
\end{abstract}


\section{Introduction}
The Richtmyer-Meshkov instability \citep{Richtmyer1960, Meshkov1969} is a hydrodynamic instability that occurs at the interface separating two fluids of different densities. It shows similarities with the Rayleigh-Taylor instability \citep{Rayleigh1883, Taylor1950} where initial perturbations at the interface grow and eventually evolve into a turbulent flow field through the transfer of potential to kinetic energy.  In the limit of an impulsive acceleration of the interface, e.g., by a shock wave, the instability is referred to as Richtmyer-Meshkov instability (RMI). In RMI, baroclinic vorticity production at the interface is caused by the misalignment of the pressure gradient ($\grad p$) associated with the shock wave and the density gradient ($\grad \rho$) of the material interface. The baroclinic vorticity production term $\left( \nabla\rho\times \nabla p \right) /\rho^2$ is the initial driving force of RMI. See \citet{Brouillette2002} and \citet{Zabusky1999} for a comprehensive review. \par
RMI occurs on enormous scales in astrophysics \citep{Arnett1989, Arnett2000, Almgren2006}, on intermediate scales in combustion \citep{Yang1993, Khokhlov1999} and on very small scales in inertial confinement fusion \citep{Lindl1992, Taccetti2005, Aglitskiy2010}. Due to the fast time scales associated with RMI, laboratory experimental measurements have difficulties to characterise quantitatively initial perturbations of the material interface and to capture the evolution of the mixing zone. General insight into the flow physics of RMI relies to a considerable extent on numerical investigations, where large-eddy simulations (LES) have become an accepted tool during the past decade.\par
\citet{Hill2006} performed a rigorous numerical investigation of RMI with re-shock. The authors used an improved version of the TCD-WENO hybrid method of \citet{Hill2004}. The method employs a switch to blend explicitly between a tuned centered-difference (TCD) stencil in smooth flow regions and a weighted essentially non-oscillatory (WENO) shock capturing stencil at discontinuities. The TCD-WENO hybrid method is used together with the stretched-vortex model \citep{Pullin2000, Kosovic2002} for explicitly modelling the subgrid interaction terms. This approach was also used by \citet{Lombardini2011} to study systematically the impact of the Atwood number for a canonical three-dimensional numerical setup, and for LES of single-shock (i.e. without re-shock) RMI \citep{Lombardini2012}. \par
\citet{Thornber2010} studied the influence of different three-dimensional broad- and narrowband multimode initial conditions on the growth rate of a turbulent multicomponent mixing zone developing from RMI. In a later study \citep{Thornber2011} the same authors presented a numerical study of a re-shocked turbulent mixing zone, and extended the theory of  Mikaelian and Youngs to predict the behaviour of a multicomponent mixing zone before and after re-shock, c.f. \citet{Mikaelian1989} and \citet{Thornber2010}. They used an implicit LES \citep{Drikakis2003, Thornber2008, Drikakis2009} approach based on a finite-volume Godunov-type method to solve the Euler equations with the same specific heat ratio for both fluids.  \par
In a recent investigation \citet{Weber2013} derived a growth-rate model for the single-shock RMI based on the net mass flux through the centre plane of the mixing zone. Here, the compressible Navier-Stokes equation were solved by a $10^{th}$-order compact difference scheme for spatial differentiation. Artificial grid dependent fluid properties, proposed by \citet{Cook2007}, were used for shock and material-interface capturing as well as for subgrid-scale modelling. \par
Grid-resolution independent statistical quantities of the single-shock RMI were presented by \citet{Tritschler2013b}. The kinetic energy spectra exhibit a Kolmogorov inertial range with $k^{-5/3}$ scaling. The spatial flux discretisation was performed in characteristic space by an adaptive central-upwind $6^{th}$-order accurate WENO scheme \citep{Hu2010} in the low-dissipation version of \citet{Hu2011}. \par
LES relies on scale separation where the energy containing large scales are resolved and the effect of non-resolved scales is modelled either explicitly or implicitly. However, turbulent mixing initiated by RMI for typical LES mainly occurs at the marginally \rev{resolved} or non-resolved scales. The interaction of non-resolved small scales with the resolved scales as well as the effect of the interaction of non-resolved scales with themselves is modelled by the employed subgrid-scale model. Moreover, discontinuities such as shock waves and material interfaces need to be captured by the numerical scheme. Due to the broad range of scales coarse-grained numerical simulations of RMI strongly rely on the resolution capabilities for the different types of subgrid scales (turbulent small scales, shocks, interfaces) of the underlying numerical scheme. \par
So far, research mainly focused on the identification and quantification of parameters that affect the evolution of Richtmyer-Meshkov unstable flows. The influence of the Atwood number \citep{Lombardini2011}, the Mach number \citep{Lombardini2012} as well as the specific initial interface perturbations \citep{Thornber2010, Grinstein2011, Schilling2010} on the temporal evolution of the instability have been investigated. Results from numerical simulations have been compared to experiments \citep{Schilling2010, Hill2006, Tritschler2013a} and theoretical models have been derived \citep{Thornber2011, Weber2013}. These investigations have assumed, based on standard arguments such as empirical resolution criteria, that the marginally and non-resolved scales have a negligible effect on the resolved scales, and therefore on the evolution of the instability. Uncertainties introduced by the numerical method, i.e., the subgrid-scale regularisation and truncation errors, have not yet been investigated systematically. There is, however, strong evidence that numerical model uncertainty can significantly affect the linear and non-linear stages of evolution, and in particular the mixing measures. In fact, it is unclear how subgrid-scale regularisation and dispersive or dissipative truncation errors can affect the resolved scales and turbulent mixing measures. \par
In the present investigation, two independently developed and essentially different numerical methods are employed to study the prediction uncertainties of RMI simulations. The first method has a dominantly dissipative truncation error at the non-resolved scales, whereas the second one exhibits a more dispersive behaviour. At the marginally resolved scales the numerical truncation error is not small and the particular character of the truncation error is essential for the implicit modelling capabilities of the method, and thus also affects the resolved-scale solution. For the purpose of investigating this effect integral and spectral mixing metrics as well as probability density functions are analysed on four computational grids with resolutions ranging from $1562~\mu m$ to $195~\mu m$. The simulations employing two different numerical methods on a very fine grid-resolution of $195~\mu m$ provide a data set with high confidence in the results. \s{Moreover, this generated data set has well-defined and thus reproducible initial and boundary conditions and thus can serve as reference for future studies. It is important to point out the significance of quantifying the influence of subgrid-scale regularisation and truncation errors in order to better understand previous works and the uncertainties associated with under-resolved simulations of RMI.} \par
\rev{We emphasise that the purpose of this study is (i) to present RMI results with a clear identification of the resolved-scale range by systematic grid refinement, (ii) to assess the physical effects of numerical subgrid-scale regularisations on the marginally resolved and on the non-resolved scale range. We do not intend to propose or improve a certain subgrid-scale model or regularisation scheme.} \par
The paper is structured as follows: The governing equations along with the employed numerical models are described in section~\ref{sec:numericalmodel}. Details about the computational domain and the exact generic initial conditions are given in section~\ref{sec:numericalsetup}. Results are presented in section~\ref{sec:results} and the key findings of the present study are discussed in section~\ref{sec:conclusion}.
\section{Numerical model} \label{sec:numericalmodel}

\subsection{Governing equations}
We solve the three-dimensional multicomponent Navier-Stokes equations
\begin{subeqnarray} \label{NS-equation1}
   \frac{\partial \rho}{\partial t} +  \nabla \cdot (\rho \boldsymbol{u}) & = & 0 \\ \label{NS-equation2}
   \frac{\partial (\rho \boldsymbol{u} ) }{\partial t} + \nabla \cdot ( \rho \boldsymbol{u u}  + p \boldsymbol{\underline{\delta}} - \underline{\boldsymbol{\tau}}) & = & 0 \\ \label{NS-equation3}
   \frac{\partial E}{\partial t} + \nabla \cdot[(E + p) \boldsymbol{u}] - \nabla \cdot (\underline{\boldsymbol{\tau}} \cdot \boldsymbol{u} - \bs{q_c} - \bs{q_d} ) & = & 0 \\ \label{NS-equation4}
   \frac{\partial \rho Y_{i} }{ \partial t } + \nabla \cdot (\rho \boldsymbol{u} Y_{i}) \s{- \nabla \cdot (\rho D_i \nabla Y_i )}\rev{+ \nabla \cdot \boldsymbol{J_i}} & = & 0  \quad .
\end{subeqnarray}
In (\ref{NS-equation1}) $\boldsymbol{u}$ is the velocity vector, $p$ is the pressure, $E$ is the total energy, $\rho$ is the mixture density, $Y_i$ is the mass fraction \rev{and $\boldsymbol{J_i}$ is the diffusive mass flux} of species $i={1,2,...K}$ with $K$ as the total number of species. The identity matrix is $\boldsymbol{\underline{\delta}}$. \par
The viscous stress tensor $\underline{\bs{\tau}}$ for a Newtonian fluid is
\begin{equation}
  \underline{\bs{\tau}} = 2 \overline{\mu} \boldsymbol{\underline{S}} + \left( \beta - 2/3 \overline{\mu} \right) \boldsymbol{\underline{\delta}} \left( \nabla\cdot\boldsymbol{u} \right) \quad ,  
\label{eq:tau}
\end{equation}
with the mixture viscosity $\overline{\mu}$ and the strain rate tensor $\boldsymbol{\underline{S}}$. According to Fourier's law we define the heat flux as
\begin{equation}
  \bs{q_c} = -\overline{\kappa} \grad T
  \label{eq:heatflux}
\end{equation}
and the interspecies diffusional heat flux \citep{Cook2009} as 
\begin{equation}
  \bs{q_d} = \sum_{i=1}^{K} h_i \bs{J_i}
  \label{eq:diffheatflux}
\end{equation}
with 
\begin{equation}
 \bs{J_i} \approx - \rho\left( D_i \grad Y_i - Y_i \sum_{j=1}^K D_j \grad Y_j \right) \quad.
 \label{eq:diffflux}
\end{equation}
$D_i$ indicates the effective binary diffusion coefficient of species $i$, and $h_i$ is the individual species enthalpy.  The equations are closed with the equation of state for an ideal gas
\begin{equation} \label{EOS_ideal}
 p(\rho e, Y_1, Y_2, ..., Y_K) = \left( \overline{\gamma} - 1 \right) \rho e \quad,
\end{equation}
where $\overline{\gamma}$ is the ratio of specific heats of the mixture and $e$ is the internal energy 
\begin{equation}
 \rho e = E - \frac{\rho}{2} \boldsymbol{u}^2 \quad.
\end{equation}
The multicomponent as well as the molecular mixing rules for $\overline{\gamma}$, $\overline{\mu}$, $D_i$, $\overline{\kappa}$ are given in appendix~\ref{ap:multicomponent mixing rules} and appendix~\ref{ap:molecular mixing rules}. \par
\subsection{Numerical methods}
\subsubsection{The Miranda simulation code}
The Miranda simulation code has been used extensively for simulating turbulent flows with high Reynolds numbers and multi-species mixing \citep{cabot:2006,cook:2004,olson:2007,olson:2011,Weber2013}. Miranda employs a $10^{th}$-order compact difference scheme \citep{lele:1992} for spatial differentiation and a five stage, $4^{th}$-order Runge-Kutta scheme \citep{kennedy:2000} for temporal integration of the compressible multicomponent Navier-Stokes equations. Full details of the numerical method are given by \citet{Cook2007} \rev{which includes an $8^{th}$-order compact filter which is applied to the conserved variables each time step and smoothly removes the top 10\% of wave numbers to ensure numerical stability.} For numerical regularisation of non-resolved steep flow gradients, artificial fluid properties are used to damp locally structures which exist on the length scales of the computational mesh. In this approach, artificial diffusion terms are added to the physical ones which appear in equations. (\ref{eq:tau}), (\ref{eq:heatflux}) and (\ref{eq:diffflux}) as
\begin{equation}
  \mu = \mu_f + \mu^{\ast} \ ,
\label{eq:shear}
\end{equation}
\begin{equation}
  \beta = \beta_f + \beta^{\ast} \ ,
\label{eq:bulk}
\end{equation}
\begin{equation}
  \kappa = \kappa_f + \kappa^{\ast} \ ,
\label{eq:cond}
\end{equation}
\begin{equation}
  D_i = D_{f,i} + D_i^{\ast} \ .
\label{eq:Diff}
\end{equation}
This LES method employing artificial fluid properties was originally proposed by \citet{Cook2007} but has been altered by replacing the $S$ (magnitude of the strain rate tensor) with $\divg\bgk{u}$ in the equation for $\beta^*$.  \citet{Mani2009} showed that this modification substantially decreases the dissipation error of the method.  Here we give the explicit formulation of the artificial terms on a Cartesian grid
\begin{equation}
  \mu^{\ast} = C_\mu \overline{\rho \left|\nabla^r S\right|} \Delta^{(r+2)} \ ,
\label{eq:mu}
\end{equation}
\begin{equation}
  \beta^{\ast} = C_{\beta} \overline{\rho \left|\nabla^r \pth{\divg\bgk{u}} \right|} \Delta^{(r+2)} \ ,
\label{eq:beta}
\end{equation}
\begin{equation}
  \kappa^{\ast} = C_{\kappa} \overline{\frac{\rho c_s}{T} \left|\nabla^r e\right|} \Delta^{(r+1)} \ ,
\label{eq:kappa}
\end{equation}
\begin{equation}
  D_i^{\ast} = C_D \overline{|\nabla^r Y_i|} \frac{\Delta^{(r+2)}}{\Delta t} 
             + C_Y \overline{\left( |Y_i| + |1-Y_i| - 1 \right)} \frac{\Delta^2}{2 \Delta t} \ ,
\label{eq:D}
\end{equation}
where $S=(\underline{\bf S}:\underline{\bf S})^{1/2}$ is the magnitude of the strain rate tensor, $\Delta=(\Delta
x \Delta y \Delta z)^{1/3}$ is the local grid spacing, $c_s$ is the sound speed and $\Delta t$ is the time-step size. The polyharmonic operator, $\nabla^r$, denotes a series of Laplacians, e.g., $r=4$ corresponds to the biharmonic operator, $\nabla^4 = \nabla^2 \nabla^2$.  The overbar $(\overline{f})$ denotes a truncated-Gaussian filter applied along each grid direction as in \citet{Cook2007} to smooth out sharp cusps introduced by the absolute value operator. \rev{In LES of RMI, $\beta^*$ acts as the shock capturing scheme. The $\mu^*$ is primarily used as a numerical stabilisation mechanism rather than an SGS model. The artificial shear viscosity is found to not be needed to maintain numerical stability in the current calculations and its inclusion has a small impact on the solution. The dissipation of the vortical motion is primarily dependent on the $8^{th}$-order compact filter.}
\subsubsection{The INCA simulation code}
The INCA simulation code is a multi-physics simulation method for single- and multicomponent turbulent flows. With respect to the objective in this paper it has been tested and validated for shock induced turbulent multi-species mixing problems at finite Reynolds numbers \citep{Tritschler2013a, Tritschler2013b, Tritschler2014a}. \par
For all simulations presented in this paper we use a discretisation scheme that employs for the hyperbolic part in (\ref{NS-equation1}) a flux projection on local characteristics. The Roe-averaged matrix required for the projection is calculated for the full multi-species system \citep{Roe1981,Larouturou1989,Fedkiw1997}. The numerical fluxes at the cell faces are reconstructed from cell averages by the adaptive central-upwind $6^{th}$-order weighted essentially non-oscillatory (WENO-CU6) scheme \citep{Hu2010} in its scale-separation formulation by \citet{Hu2011}.\par
The fundamental idea of the WENO-CU6 scheme is to use a non-dissipative $6^{th}$-order central stencil in smooth flow regions and a non-linear convex combination of $3^{rd}$-order stencils in regions with steep gradients. The reconstructed numerical flux at the cell boundaries is computed from
\begin{equation}
 \hat{f}_{i+1/2} = \sum^3_{k=0} \omega_k \hat{f}_{k,i+1/2} \quad,
\end{equation}
where $\omega_k$ is the weight assigned to stencil $k$ with the $2^{nd}$-degree reconstruction polynomial approximation for $\hat{f}_{k,i+1/2}$. In the WENO-CU6 framework the weights $\omega_k$ are given by
\begin{equation} \label{eq:eightingstrategy}
 \omega_k = \frac{\alpha_k}{\sum^3_{k=0} \alpha_k} \mbox{,} \quad \alpha_k = d_k \left( C+\frac{\tau_6}{\beta_k + \epsilon } \right)^q \quad ,
\end{equation}
with $\epsilon$ being a small positive number $\epsilon = 10^{-40}$. The optimal weights $d_k$ are defined such that the method recovers the $6^{th}$-order central scheme in smooth flow regions. The constant parameters in (\ref{eq:eightingstrategy}) are set to $C = 1000$ and $q=4$, see \citet{Hu2011} $\tau_6$ is a reference smoothness indicator that is calculated from a linear combination of the other smoothness measures $\beta_k$ with 
\begin{equation}
 \tau_6 = \beta_6 - \frac{1}{6}( \beta_0 + \beta_2 + 4 \beta_1 )
\end{equation}
and
\begin{equation}\label{smoothness_indicators}
 \beta_k = \sum^2_{j=1} \Delta x^{2j-1} \int^{x_{i+1/2}}_{x-1/2} \left( \frac{d^j}{dx^j} \hat{f}_k(x) \right)^2 dx \quad.
\end{equation}
$\beta_6$ is also calculated from (\ref{smoothness_indicators}) but with the $5^{th}$-degree reconstruction polynomial approximation of the flux which gives the 6-point stencil for the $6^{th}$-order interpolation. \par
After reconstruction of the numerical fluxes at the cell boundaries the fluxes are projected back onto the physical field. A local switch to a Lax-Friedrichs flux is used as entropy fix, see, e.g., \citet{Toro1999}. A positivity-preserving flux limiter \citep{Hu2013} is employed in regions with low pressure or density, maintaining the overall accuracy of the $6^{th}$-order WENO scheme. It has been verified that the flux limiter has negligible effect on the results, and avoids excessively small time step sizes. Temporal integration is performed by a $3^{rd}$-order total variation diminishing 
Runge-Kutta scheme \citep{Gottlieb1998}.
\section{Numerical setup}\label{sec:numericalsetup}
\subsection{Computational domain}
\begin{figure}
  \centering
    \includegraphics[width=0.6\textwidth]{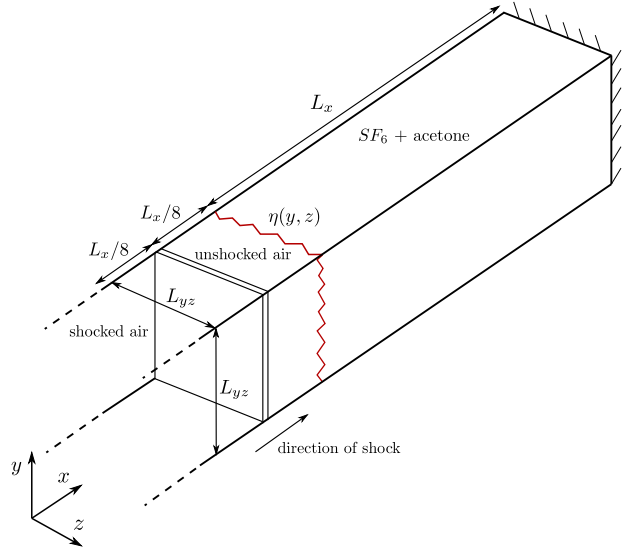}
  \caption{~Schematic of the square shock tube and dimensions of the computational domain for the simulations.~}
  \label{figure:ECC_schematic}
\end{figure}

We consider a shock tube with constant square cross section. The fine-grid domain extends in the $y$- and $z$-direction symmetrically from $-L_{yz}/2$ to $L_{yz}/2$ and from $-L_x/4$ to $L_x$ in the x-direction. An inflow boundary condition is imposed far away from the fine-grid domain in order to avoid shock reflections. To reduce computational costs a hyperbolic mesh stretching is applied between the inflow boundary and $-L_x/4$. $L_x$ is set to $0.4~m$, and $L_{yz} = L_x/4$. At the boundaries normal to the $y$- and $z$-direction, periodic boundary conditions are imposed and an adiabatic-wall boundary at the end of the shock tube at $x = L_x$ is used. A schematic of the computational domain is shown in figure~\ref{figure:ECC_schematic}. \par
The fine-grid domain is discretised by four different homogeneous Cartesian grids with \s{planar resolutions of} $64$, $128$, $256$ and $512$ cells \rev{in the y- and z-direction and $320$, $640$, $1280$ and $2560$ cells in the x-direction} resulting in cubic cells of size $1562~\mu m \lesssim \Delta_{xyz} \lesssim 195~\mu m$. The total number of cells in the fine-grid domain amounts to $\approx 1.3 \cdot 10^6$ for the coarsest resolution and to $\approx 670 \cdot 10^6$ for the finest resolution.
\subsection{Initial conditions}
We consider air as mixture of nitrogen ($N_2$) and oxygen ($O_2$) with (in terms of volume fraction) $X_{N_2} = 0.79$ and $X_{O_2} = 0.21$. The equivalent mass fractions on the air side give $Y_{N_2} = 0.767$ and $Y_{O_2} = 0.233$\rev{, i.e., $Y_{air} = Y_{N_2} + Y_{O_2}$}. The heavy gas \s{side} is modelled as a mixture of $SF_6$ and acetone ($Ac$) with mass fractions $Y_{SF_6} = 0.8$ and $Y_{Ac} = 0.2$\rev{, i.e., $Y_{HG} = Y_{SF_6} + Y_{Ac}$}. The material interface between light (air) and heavy gas is accelerated by a $\Ma = 1.5$ shock wave that is initialised at $x= -L_x/8$ propagating in the positive $x$-direction. The pre-shock state is defined by the stagnation condition $p_0 = 23000~Pa$ and $T_0 = 298~K$. The corresponding post-shock thermodynamic state is obtained from the Rankine-Hugoniot conditions
\begin{subeqnarray}
 \rho'_{air} & = & \rho_{air} \frac{( \gamma_{air} + 1 ) \Ma^2}{ 2 + ( \gamma_{air} - 1 ) \Ma^2 } \\
 u'_{air}    & = & \Ma~c_{air} \left( 1 - \frac{\rho_{air}}{\rho'_{air}} \right) \\
 p'_{air}    & = & p_0 \left( 1 + 2  \frac{ \gamma_{air}}{ \gamma_{air} + 1 } \left( \Ma^2 - 1 \right)  \right)
\end{subeqnarray}
with $c_{air} = \sqrt{ \gamma_{air} p_0 / \rho_{air} }$. The initial data of the post-shock state of the light gas as well as the pre-shock state of the light and heavy gas are given in table \ref{table:initialvalue}.\par
\begin{table}
  \centering
    \begin{tabular}{ccccc} 
    \hline
    \bf{Quantity}      & \bf{Post-shock}       & \bf{Pre-shock light-gas side}   & \bf{Pre-shock heavy-gas side}  \\
    $\rho~[kg/m^3]$    & $0.49869$             & $0.26784$            & $1.04057$              \\ 
    $U~[m/s]$          & $240.795$             & $0$                  & $0$                    \\ 
    $p~[Pa]$           & $56541.7$             & $23000$              & $23000$                \\ 
    $T~[K]$            & $393.424$             & $298$                & $298$                  \\ 
    $D_{N_2}~[m^2/s]$  & $5.919\cdot10^{-05}$  & $8.981\cdot10^{-05}$ & -                      \\ 
    $D_{O_2}~[m^2/s]$  & $5.919\cdot10^{-05} $ & $8.981\cdot10^{-05}$ & -                      \\ 
    $D_{SF_6}~[m^2/s]$ & -                     & -                    & $1.846\cdot10^{-05}$   \\ 
    $D_{Ac}~[m^2/s]$   & -                     & -                    & $1.846\cdot10^{-05}$   \\ 
    $\overline{\mu}~[Pa s]$& $ 2.234\cdot10^{-05}$ & $1.826\cdot10^{-05} $ & $1.328\cdot10^{-05}$ \\ 
    $\overline{ c_p}~[J/(kg K)]$& $1008.35 $   & $1008.35 $           & $815.89$              \\ 
    \end{tabular}
    \caption{~~Initial values of the post-shock state and the pre-shock states of the light and heavy-gas side.~~}
  \label{table:initialvalue}
\end{table}
\citet{Tritschler2013b} introduced a generic initial perturbation of the material interface that resembles a stochastic random perturbation being, however, deterministic and thus exactly reproducible for different simulation runs. This multimode perturbation is given by the following function
\begin{eqnarray} \label{eq:initial}
 \eta(y,z) & = & a_1 \sin\left( k_0 y \right) \sin\left( k_0 z \right) \nonumber \\
           & + & a_2 \sum_{n=1}^{13}\sum_{m=3}^{15} a_{n,m} \sin\left( k_n y + \phi_n \right) \sin\left( k_m z + \chi_m \right) 
\end{eqnarray}
with the constant amplitudes $a_1 = - 0.0025~m$ and $a_2 = 0.00025~m$ and wavenumbers $k_0 = 10 \pi / L_{yz}$, $k_n = 2 \pi n / L_{yz}$ and 
$k_m = 2 \pi m / L_{yz}$. The \s{random} amplitudes $a_{n,m}$ and the phase shifts $\phi_n$ and $\chi_m$ are given by
\begin{subeqnarray}
  a_{n,m} & = & \sin(n m)/2 \\
  \phi_n  & = & \tan(n) \\
  \chi_m  & = & \tan(m) \quad .
\end{subeqnarray}
For \s{verifying grid convergence}\rev{facilitating a grid sensitivity study} we impose an initial length scale by prescribing a finite initial interface thickness in the mass fraction field as
\begin{equation}
  \psi(x,y,z) = \frac{1}{2} \left( 1 + \tanh\left( \frac{ x - \eta(y,z) }{L_\rho } \right) \right)
\end{equation}
with $L_\rho = 0.01~m$ being the characteristic initial thickness.  The individual species mass fractions are \s{then} set as
\begin{eqnarray}
  Y_{SF_6} & = & 0.8 \ \psi, \quad \quad \quad \quad
  Y_{Ac} = 0.2 \ \psi ,\nonumber\\
  Y_{N_2} & = & 0.767 \ (1 - \psi) ,\quad
  Y_{O_2} = 0.233 \ (1 - \psi) . 
\end{eqnarray}
The material interface is initialised at $x - \eta(y,z) = 0~m$. Combined with the multicomponent and molecular mixing rules given in appendix~\ref{ap:multicomponent mixing rules} and appendix~\ref{ap:molecular mixing rules}, the flow field is \s{then} fully defined at $t=0$. \par
\begin{figure}
  \centering
    \includegraphics[width=0.6\textwidth]{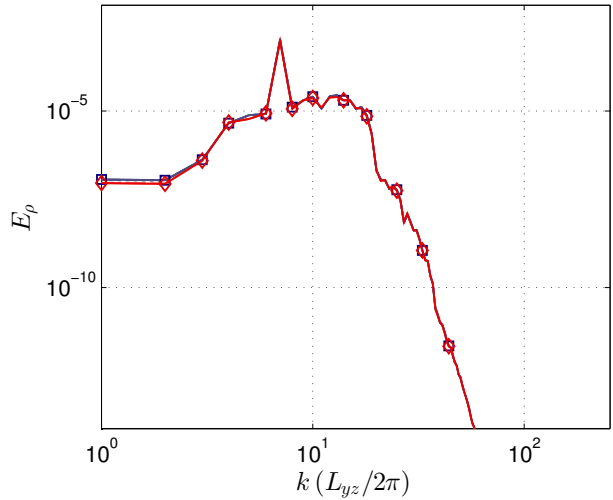}
  \caption{~Initial power spectra of density from Miranda (blue/dark grey) and INCA (red/light grey). The different resolutions are represented as dotted line ($64$), dashed line ($128$), solid line ($256$) and solid line with open squares for Miranda and open diamonds for INCA ($512$).~}
  \label{figure:initial}
\end{figure}

 
\rev{Figure~\ref{figure:initial} shows the initial condition in terms of the power spectrum of density for Miranda and INCA at all grid resolutions. The initial perturbation given in (\ref{eq:initial}) and shown in figure~\ref{figure:initial} has been designed with the objective to obtain a reproducible and representative data set. Nevertheless, we cannot exclude that some of the observations presented in this paper do not apply to very different initial perturbations.}
\section{Results}\label{sec:results}
\begin{figure}
 \centering
 \includegraphics[width=.6\textwidth]{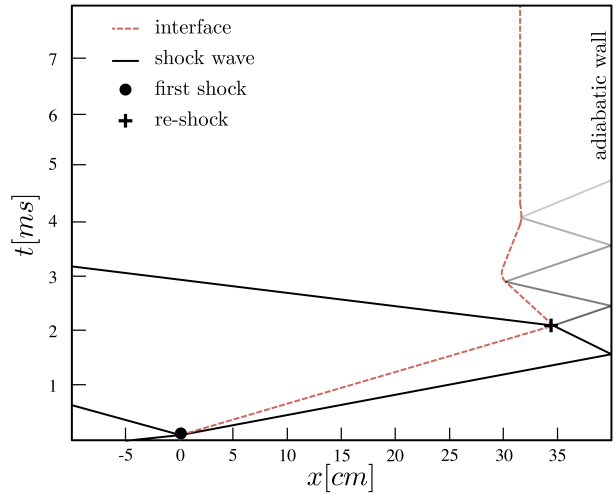}
 \caption{\label{figure:XT}
Space-time ($x-t$) diagram depicting the propagation of the shock wave and interface during the simulation. The effect of the shock wave on the interface location is attenuated with each subsequent reflection. }
\end{figure}

For exploring the effect of the finite truncation error arising from grid resolution and numerical method, four meshes were used to compute the temporal evolution of RMI with both Miranda and INCA. \s{Isotropic grid spacing is used with four different cross-flow-plane resolutions of $64^2$, $128^2$, $256^2$ and $512^2$ cells.} The simulation reaches $t=6.0\ ms$ which is well beyond the occurrence of re-shock \rev{at $t \approx 2~ms$}. At this stage, the effects of reflected shock waves and expansion waves on the shock location has become small \rev{as the shock wave are attenuated with each subsequent reflection.} \s{as can be seen in figure~\ref{figure:XT}.}\rev{The space-time ($x-t$) diagram shown in figure~\ref{figure:XT} depicts the propagation of the shock wave and interface during the simulation.} \par  
The initial conditions described in the previous section are entirely deterministic and \s{as having}\rev{due to their} band-limited representations are \s{also} identically \s{exact} imposed at the different grid resolutions and for the two numerical methods. Therefore, the obtained results exhibit uncertainties only due to the numerical method and due to grid resolution, but exclude initial-data uncertainties. \s{Moreover, as resolution is increased results of both numerical models converge towards a unique solution.}\par
\begin{figure}
 \centering
 \includegraphics[width=1\textwidth]{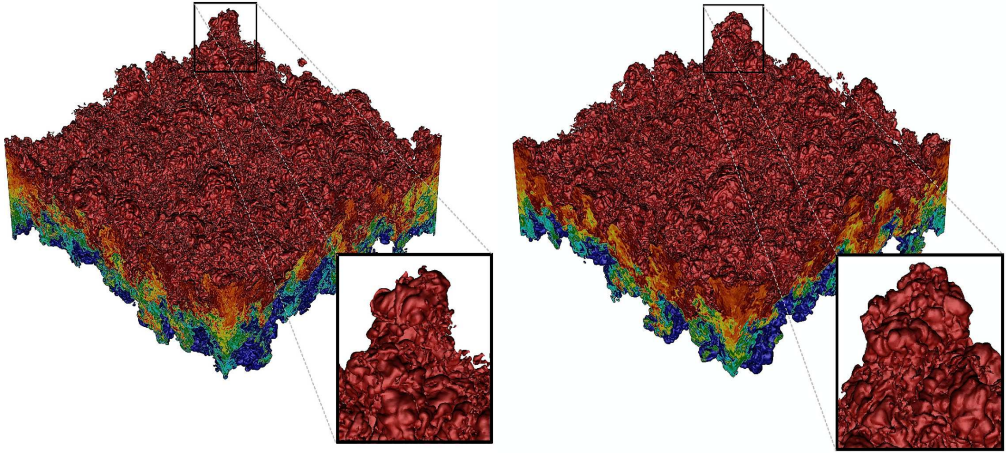}
 \caption{\label{figure:3d}
Three-dimensional contour plots of species mass fraction of the heavy gas from Miranda (left) and INCA (right) data. Data are from the finest grid at $t=2.5~ms$ that show contours of the heavy gas mass fraction $Y_{HG}$ from 0.1 (blue) to 0.9 (red).  Note that although some large scale features remain consistent between codes, small and intermediate scales are quite different at this stage. }
\end{figure}

For illustration we show the three-dimensional contour plots of species mass fraction of the heavy gas $Y_{HG}$ obtained with Miranda and INCA, respectively, in figure~\ref{figure:3d}. Similarities at the large scales are clearly visible after re-shock, but also differences exist at the fine scales, more clearly visible from the inset.
\subsection{Integral quantities}
Integral measures of the mixing zone are presented here for both numerical models and all resolutions. Often, these time dependent integral measures are the only metrics available for comparison with experiment and are therefore of primary importance for validation. \par
\begin{figure}
  \centering
    \includegraphics[width=0.8\textwidth]{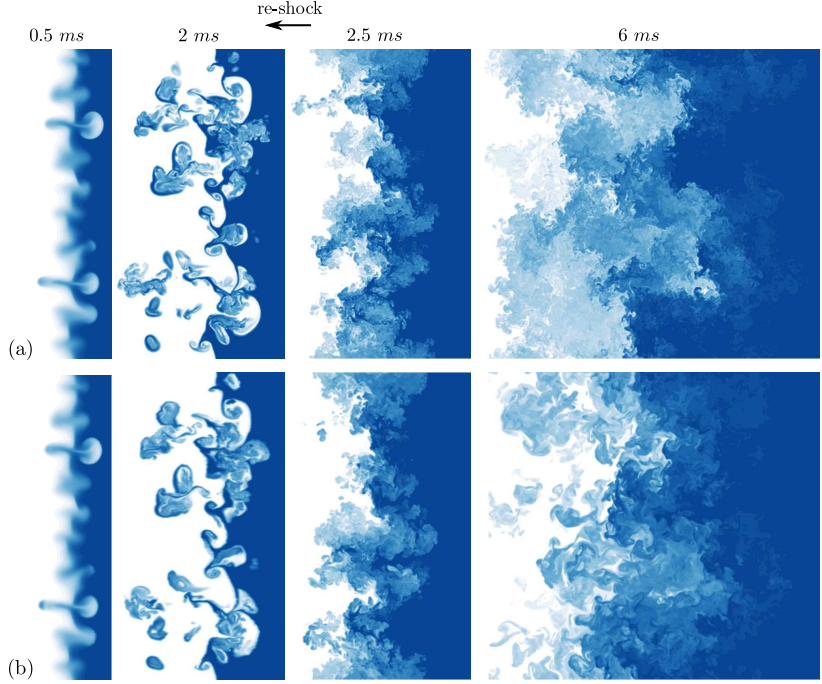}
  \caption{~Colour coded plots of species mass fraction of $SF_6$ gas from Miranda (a) and INCA (b) at various times where as data are taken from the finest grid. The contours range from 0.05 (white) 
to 0.75 (blue).~}
  \label{figure:contour}
\end{figure}

Figure~\ref{figure:contour} shows the transition process predicted by the reference grid with a \s{planar} resolution of $512$ cells \rev{in the transverse directions}. The numerical challenge, prior to re-shock, is to predict the large-scale non-linear entrainment and the associated interface steepening. The interface eventually becomes under-resolved when its thickness reaches the resolution limit of the numerical scheme and further steepening is prevented by numerical diffusion. The equilibrium between interface steepening and numerical diffusion occurs later in time as the grid is refined. \rev{The accurate prediction of the interface steepening phenomenon is one of the main challenges in modelling pre-transitional RMI where large-scale flow structures are still regular.} \s{In the following our results will verify that the accurate prediction of the interface steepening phenomenon is one of the main challenges in modelling pre-transitional RMI where large-scale flow structures are still regular.} This is because the numerical model largely determines the time when mixing transition occurs. In nature, mixing transition is due to the presence of small-scale perturbations whereas in numerical simulation, the transition is triggered by back-scatter from the under-resolved scales as predicted by the particular numerical model. Hence, details of mixing transition of the material interface evolve differently for the two codes. \par
\begin{figure}
  \centering
    \includegraphics[width=1\textwidth]{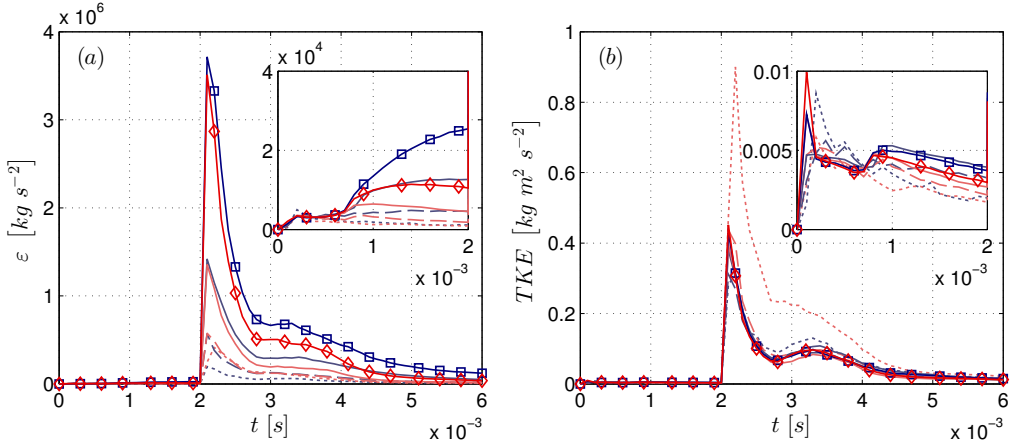}
  \caption{~Enstrophy $\varepsilon$ and turbulence kinetic energy ($\tke$) from Miranda (blue/dark grey) and INCA (red/light grey). The different resolutions are represented as dotted line ($64$), 
dashed line ($128$), solid line ($256$) and solid line with open squares for Miranda and open diamonds for INCA ($512$).~}
  \label{figure:TKE_Enstrophy}
\end{figure}

Nevertheless, similarities before re-shock are striking and large-scale similarities in the resolved wavenumber range even persist throughout the entire simulation time. Following re-shock the large  interfacial scales break down into smaller scales and develop a turbulent mixing zone as can be seen in figure~\ref{figure:3d} and figure~\ref{figure:contour}. By visual inspection of figure~\ref{figure:contour} one finds that the post-re-shock turbulent structures are very similar, whereas the long term evolution of the small scales appears to be different between the codes. Differences in the observed flow field at $t=6~ms$ may indicate slightly different effective Reynolds numbers for the two numerical methods and therefore \rev{they} also \s{show}\rev{exhibit} different decay rates of enstrophy \citep{Dimotakis2000, Lombardini2012}, as can be seen in figure~\ref{figure:TKE_Enstrophy} after re-shock.\par
\begin{figure}
  \centering
    \includegraphics[width=1\textwidth]{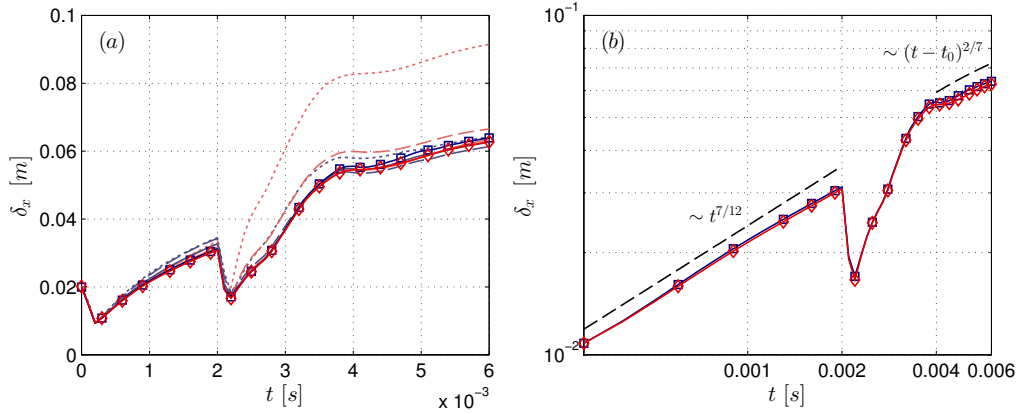}
  \caption{~Time evolution of the mixing zone width from Miranda (blue/dark grey) and INCA (red/light grey). The different resolutions are represented as dotted line ($64$), 
dashed line ($128$), solid line ($256$) and solid line with open squares for Miranda and open diamonds for INCA ($512$).~}
  \label{figure:MLW_combined}
\end{figure}

The mixing width $\delta_x$ is a length scale that approximates the large-scale temporal evolution of the turbulent mixing zone. It is defined as an integral measure by
\begin{equation}
 \delta_x(t) = \int_{-\infty}^{\infty} 4 \phi\left( 1 - \phi \right) dx \ , \ \ \mbox{with} \quad  \phi(x,t) = \left< Y_{SF_6} + Y_{Ac} \right>_{yz} \ ,
\end{equation}
where $\left< \cdot \right>_{yz}$ denotes the ensemble average in the cross-stream yz-plane. For a quantity $\varphi$ it is defined by 
\begin{equation}
 \left< \varphi \right>_{yz}(x,t) = \frac{1}{\mathcal{A}} \iint \varphi(x,y,z,t) dy dz \ , \ \  \mbox{with} \quad \mathcal{A} = \iint dy dz \ .
\end{equation}
The mixing width plotted in figure~\ref{figure:MLW_combined}(a) shows that data from both numerical methods converge to a single solution throughout the entire simulation time. Furthermore, it is observed that even with very-high-order models a minimum resolution of $\sim 400~\mu m$ appears to be necessary for an accurate prediction of the mixing-zone-width. As will be shown later, coarser grids not only tend to overpredict the growth of the mixing zone but also molecular mixing. \par
Figure~\ref{figure:MLW_combined}(b) shows the mixing zone width time evolution on a log-log scale. The (bubble) growth rate model of \citet{Zhou2001} predicts accurately the pre-re-shock mixing zone growth rate that is consistently recovered by both numerical methods as $\sim t^{7/12}$. However, this is\rev{, according to \citet{Zhou2001},} the growth rate that is associated with turbulence of Batchelor type \citep{Batchelor1956} with $E(k) \sim k^4$ as $k \rightarrow 0$. The kinetic energy spectra in the present investigation are of Saffman type \citep{Saffman1967a, Saffman1967b} with $E(k) \sim k^2$ as $k \rightarrow 0$ \citep{Tritschler2013b} for which \citet{Zhou2001} predicts a growth that scales with $\sim t^{5/8}$. \rev{The present growth rates are also in good agreement with the experimental and numerical results of \citet{Dimonte1995} with $\sim t^\beta$ and $\beta = 0.6 \pm 0.1$ and their model predictions $\sim \left( t-t_i\right)^{1/2}$, where $t_i$ accounts for the time the shock needs to traverse the interface.} \s{Note that direct comparison with theory is difficult, as the mixing zone has not yet reached self-similar evolution where the growth rate is independent of the initial condition.}\rev{As the mixing zone has not yet reached self-similar evolution, the initial growth rate depends on the specific initial conditions.}\par
\citet{Llor2006} found that the self-similar growth rate of the energy-containing eddies, i.e., the integral length scale, for incompressible RMI at vanishing Atwood number should scale as $\delta_x \sim t^{1-n/2}$ with $2/7 \le 1-n/2 \le 1/3$\rev{, if the turbulence kinetic energy decays as $\sim t^{-n}$}. These growth rates slightly differ from the growth rate prediction for homogeneous isotropic turbulence $1/3 \le 1-n/2 \le 2/5$ by the same author. \rev{The predictions of \citet{Llor2006}, however, are at odds with Kolmogorov's classical decay law  \citep	{Kolmogorov1941} for turbulence kinetic energy $\sim t^{-10/7}$ and more recent investigations of decaying isotropic turbulence by \citet{Ishida2006} and \citet{Wilczek2011}, which found Kolmogorov's decay law to hold if the Loitsyansky integral is constant and if the Taylor-scale Reynolds number exceeds $Re_\lambda \ge 100$.}
Based on Rayleigh-Taylor experiments driven by either sustained or impulsive acceleration at various Atwood numbers, \citet{Dimonte2000} found scaling laws for the bubble and spike growth rate. For the present density ratio the exponents become $1-n_B/2 \approx 0.25 \pm 0.05$ for the former and $0.25 \le 1-n_S/2 \le 0.43$ for the latter. The late-time mixing zone growth rate is therefore expected to correlate with the spike growth rate. The late-time growth rate prediction of the present work is $\sim (t-t_0)^{2/7}$, i.e., $1-n/2 = 2/7$, once the turbulent mixing zone is fully established. $t_0$ is a virtual time origin set to $t_0 = 2~ms$. This is consistent with the mixing zone width growth rate predictions of \citet{Llor2006} and the late time growth rate predictions of \citet{Dimonte2000}, but underestimates the predictions of \citet{Zhou2001}, \s{who predicts the turbulent mixing zone width to scale as}\rev{with a scaling of} $t^{1-n/2}$ with $0.35 \le 1-n/2 \le 0.45$ long after re-shock, once the non-linear time scale has become the dominant time scale. \rev{In the numerical investigation of} \citet{Lombardini2012} \rev{the authors} found the mixing zone width to grow as $0.2 \le 1-n/2 \le 0.33$ \s{in their numerical investigation}. \rev{Before re-shock the infrared part of the kinetic energy spectrum, see figure~\ref{figure:E_kin}, exhibits a $k^2$ range, for which a post-re-shock growth rate of $\sim t^{2/7}$ is predicted by the model of \citet{Youngs2004}, which is in good agreement with the present data.} \par
The definition of the molecular mixing fraction $\Theta$ \citep{Youngs1994, Youngs1991} is given as
\begin{equation}
 \Theta(t) = \frac{ \int_{-\infty}^{\infty}{ \left< X_{air} X_{HG} \right>_{yz} dx} } 
                  { \int_{-\infty}^{\infty}{ \left<X_{air}\right>_{yz}  \mbox{ } \left< X_{HG} \right>_{yz} dx} }
\end{equation}
and quantifies the amount of mixed fluid within the mixing zone. It can be interpreted as the ratio of molecular mixing to large-scale entrainment by convective motion. \par
\begin{figure}
  \centering
    \includegraphics[width=1\textwidth]{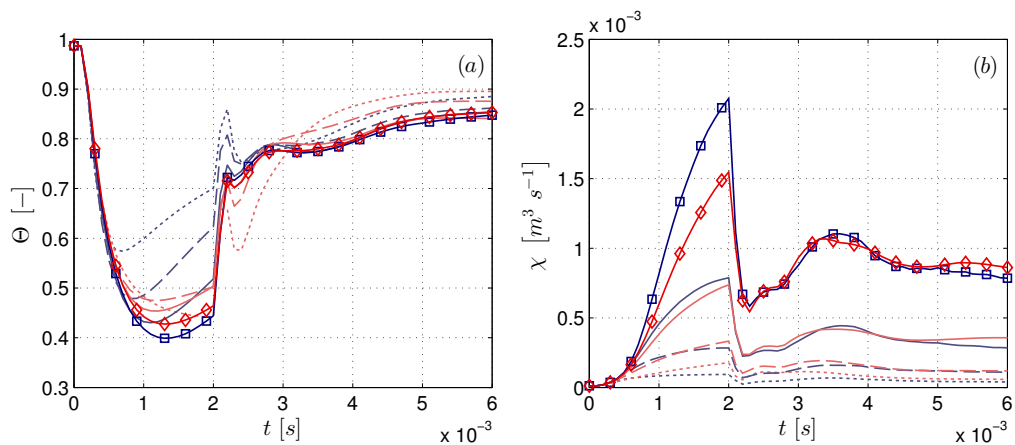}
  \caption{~Molecular mixing fraction $\Theta$ and scalar dissipation rate $\chi$ from Miranda (blue/dark grey) and INCA (red/light grey). The different resolutions are represented as dotted line ($64$), dashed line ($128$), solid line ($256$) and solid line with open squares for Miranda and open diamonds for INCA ($512$).~}
  \label{figure:MMF_TMR}
\end{figure}

As bubbles of light air and spikes of heavy gas begin to interfuse, the initially mixed interface between the fluids steepens and the fluids become more segregated on the molecular level, see figure~\ref{figure:MMF_TMR}(a). The molecular mixing fraction reaches its minimum at $t \approx 1.3~ms$ before Kelvin-Helmholtz instabilities lead to an increase of molecular mixing. \rev{The onset of secondary instabilities is very sensitive to the numerical method as the numerical scheme determines how sharp the material interface can be represented or whether numerical diffusion or dispersion effects lead to an early mixing transition.}\par
After re-shock molecular mixing is strongly enhanced and reaches its maximum of $\Theta \approx 0.85$ by the end of the simulation. This finding is consistent with \citet{Lombardini2012} who also found an asymptotic late-time mixing behaviour with $\Theta \approx 0.85$ independent of the shock Mach number but without re-shock. The asymptotic limit is already \s{correctly}\rev{accurately} calculated on grid-resolutions of $\sim 400~\mu m$. As the second shock wave compresses the mixing zone, the instability becomes less entrained yet equally diffused (at least in the y- and z-directions) and therefore causes a steep rise in $\Theta$. A gradual increase of the mixing fraction after the steep rise occurs as the mixing zone becomes more homogeneously distributed \citep{Thornber2011} due to turbulent motion.\par
The temporal evolution of the scalar dissipation rate is plotted in figure~\ref{figure:MMF_TMR}(b) and is derived from the advection-diffusion equation for a scalar. The instantaneous scalar dissipation rate of the three-dimensional RMI is estimated from the $SF_6$ concentration field as
\begin{equation}
 \chi(t) = \int_{-\infty}^{\infty}{ D_{SF6} \nabla Y_{SF6} \cdot \nabla Y_{SF6}~ dx dy dz } \ ,
 \label{eq:TMR}
\end{equation}
which quantifies the rate at which mixing occurs. For consistency of post-processing, a $2^{nd}$-order central-difference scheme has been used for the calculation of the spatial derivatives in  (\ref{eq:TMR}) and (\ref{eq:Enst}) for all simulation data sets. \rev{Note that the order of the finite-difference scheme with which the gradients in (\ref{eq:TMR}) and (\ref{eq:Enst}) are approximated affect their results. }\par
The variation of the scalar dissipation rate with grid resolution before re-shock is largely due to the under-resolved material interface and the onset of mixing transition. Mixing is strongly enhanced after the second shock-interface interaction, but the mixing zone is also confined to a much smaller region which results in a decrease of the integral $\chi$. Also, $\chi$ only represents the resolved part of the dissipation rate and therefore certainly underestimates the true value.\par
The turbulence kinetic energy ($\tke$) and the enstrophy ($\varepsilon$) are integrated over cross-flow planes in the mixing zone that satisfy 
\begin{equation} \label{eq:imz}
 4 \phi \left[ 1 - \phi \right] \ge 0.9 \quad .
\end{equation}
This region is referred to as the inner mixing zone ($\imz$) in the following.\par
Baroclinic vorticity is deposited at the material interface during shock passage. The amount of generated vorticity scales directly with the pressure gradient of the shock wave and the density gradient of the material interface. The enstrophy is calculated by
\begin{equation}
  \varepsilon(t) = \s{\int_{IMZ}{ \rho \left| \boldsymbol{\omega} \right|^2 dx dy dz } =} \int_{IMZ}{ \rho \left( \omega_i \omega_i \right) dx dy dz } \quad,
  \label{eq:Enst}
\end{equation}
where $\omega_i$ is the vorticity. \par
As can be seen from figure~\ref{figure:TKE_Enstrophy}, the enstrophy also exhibits a strong grid dependency. Fully grid-converged results are only obtained for times up to $t \approx 0.7~ms$. As the interface steepens due to strain and shear the effective interface thickness is determined by numerical diffusion which appears to occur at $t \approx 0.7~ms$. This is consistent with the evolution of $\Theta$ shown in figure~\ref{figure:MMF_TMR}(a). \rev{Following \citet{Hahn2011} and \citet{Youngs2007} integration of enstrophy with a theoretical scaling of $k^{1/3}$ up to the cut-off wavenumber yields a proportionality between enstrophy and grid resolution as $\varepsilon \propto \Delta_{xyz}^{-4/3}$. From this follows an increase of enstrophy by a factor of about $2.5$ from one grid resolution to the next finer, which is in good agreement with the present data.}\par
The amount of turbulence kinetic energy created by the impulsive acceleration of the interface is calculated as
\begin{equation}
 \tke(t) = \int_{IMZ}{ K~dx dy dz} \quad \mbox{,with} \quad  K(x,y,z,t) = \frac{\rho}{2} u''_i u''_i \ .
\end{equation}
The fluctuating part $\varphi''$ of a quantity $\varphi$ is calculated from 
\begin{equation}
\varphi'' = \varphi - \overline{\varphi}  \quad \mbox{,with} \quad \overline{\varphi} = \left<\rho \varphi \right>_{yz} / \left<\rho \right>_{yz} ,
\end{equation}
where $\overline{\varphi}$ is the Favre-average of $\varphi$. \par
Grid-converged turbulence kinetic energy $\tke$ is obtained on grids with a minimum resolution of $\sim 400~\mu m$, see figure~\ref{figure:TKE_Enstrophy}. This is consistent with the convergence rate of the mixing zone width. The total $\tke$ deposited in the $\imz$ by the first shock-interface interaction can be seen in the inset of figure~\ref{figure:TKE_Enstrophy}. The re-shock occurring at $t \approx 2~ms$ deposits approximately $40$ times more $\tke$ than the initial shock wave. \citet{Hill2006} found a similar relative increase by the re-shock at the same shock Mach number. A significant decay in energy occurs immediately following re-shock.  The material interface \s{then} interacts with the first expansion fan, see figure~\ref{figure:XT}, and results in a further increase in $\tke$ between $3~ms - 3.5~ms$. The amount of energy deposited by the first expansion wave, however, is much weaker than that deposited by the reflected shock wave. \citet{Grinstein2011} and \citep{Hill2006} found the amplification of $\tke$ by the first rarefaction to be much stronger than for our data. Such differences are not surprising as \citet{Grinstein2011} reported a strong dependence of energy deposition on the respective initial interface perturbations. After the first expansion wave has interacted with the interface, $\tke$ decays slowly and the pressure gradients associated with the subsequent rarefactions are too shallow to generate any further noticeable increase in $\tke$. \par
\begin{figure}
  \centering
    \includegraphics[width=0.5\textwidth]{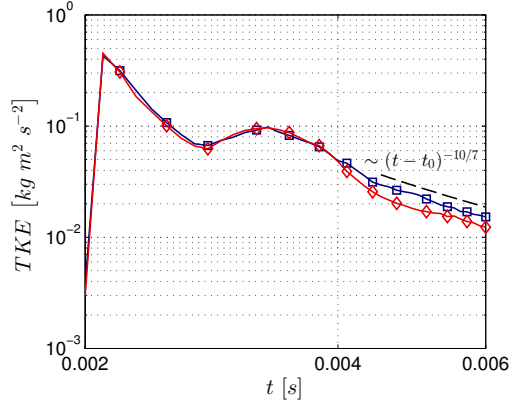}
  \caption{~Log-log representation of $\tke$ from Miranda (squares) and INCA (diamonds) taken from the finest grid ($512$).~}
  \label{figure:TKE_loglog}
\end{figure}
\citet{Lombardini2012} found the decay rate of $\tke$ to be larger than $\sim t^{-6/5}$ approaching $\sim t^{-10/7}$. In our data, the late time $\tke$ decay is also approximately $\sim (t-t_0)^{-10/7}$ with $t_0 = 2~ms$ being the virtual time origin\rev{, see figure~\ref{figure:TKE_loglog}}. This scaling would be characteristic for Batchelor-type turbulence \citep{Batchelor1956} \rev{with a constant Loitsyansky integral \citep{Kolmogorov1941, Ishida2006}} in contrast to $\sim t^{-6/5}$ typical for turbulence of Saffman-type \citep{Saffman1967a, Saffman1967b}. \rev{The similar $\tke$ decay rates in homogeneous isotropic turbulence and \citet{Lombardini2012} and the present data suggest that the dissipation mechanism in RMI is quite similar to that in homogeneous isotropic turbulence.}\par
\s{It is worth noting that after some time the turbulent mixing zone is expected to reach a self-similar quasi-isotropic state where the mean turbulence kinetic energy scales with $\tke \propto t^{-n} $ and the integral length scale, which can be related to the mixing zone width, as $\delta_x \propto t^{1-n/2}$.} \rev{In the limit of a self-similar quasi-isotropic state the temporal evolution of the integral length scale $\delta_x$ is related to the evolution of $\tke$ in the mixing zone. From $\tke \propto t^{-n}$ the growth rate of the integral scale follows as $\delta_x \propto t^{1-n/2}$.} \citet{Llor2006} derived a maximum decay rate of turbulence kinetic energy $\sim t^{-10/7}$ that corresponds to a growth rate scaling of the energy-containing eddies of $\delta_x \sim t^{2/7}$. These predictions are in excellent agreement with the growth rate predictions of the mixing zone width of the present investigation, see figure~\ref{figure:MLW_combined}, and the decay rate of $\tke$, see figure~\ref{figure:TKE_loglog}. \par
\rev{The scalings indicated for the growth rate of the mixing zone and the decay rate of $\tke$ in figures \ref{figure:MLW_combined} and \ref{figure:TKE_loglog} were not fitted in a strict sense. They merely serve as reference for comparison with incompressible isotropic decaying turbulence. The narrow data range of only $\approx 2~ms$ after re-shock for which the flow exhibits a self-similar regime precludes any precise estimates for decay and growth rate laws.} \par
\begin{figure}
  \centering
    \includegraphics[width=1\textwidth]{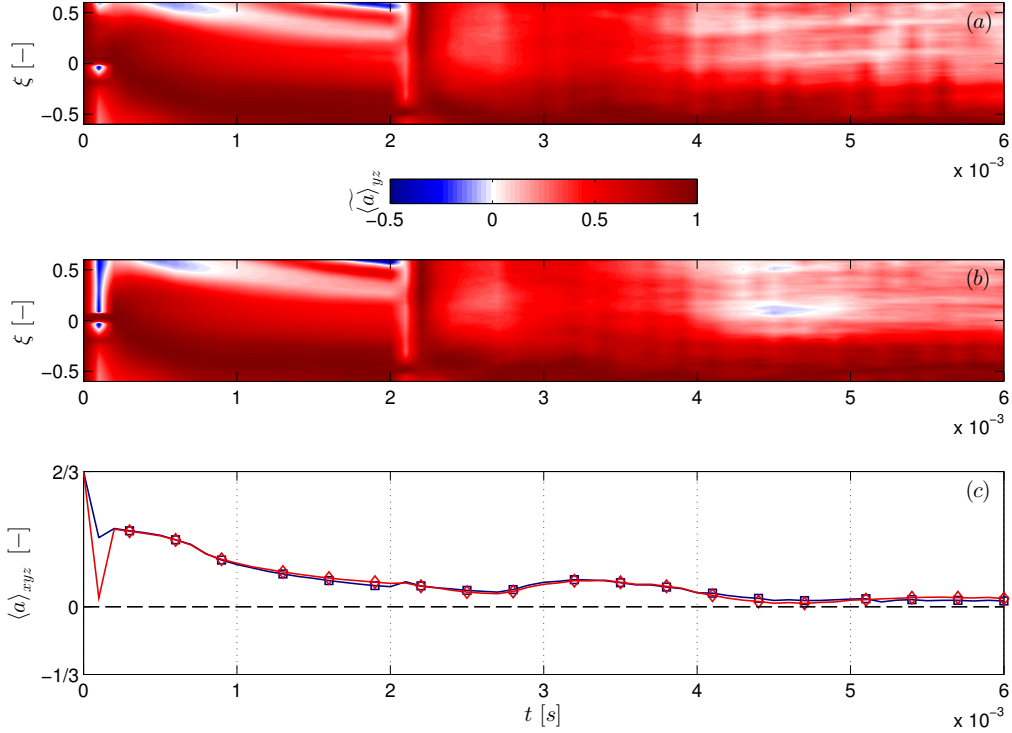}
  \caption{~Anisotropy $\left< a \right>_{yz}$ as a function of the dimensionless mixing zone coordinate $\xi$ and time from Miranda (a) and INCA (b). The volume averaged anisotropy $\left< a \right>_{xyz}$ of the inner mixing zone is shown in (c) from Miranda (squares) and INCA (diamonds). All data are taken from the finest grid ($512$).~}
  \label{figure:2D_iso}
\end{figure}

\rev{\subsection{Anisotropy and inhomogeneity of the mixing zone}}
\rev{In the following the anisotropy in the mixing zone is investigated. We define the local anisotropy as}
\begin{equation}
 \rev{ a(x,y,z,t) = \frac{|u''|}{|u''|+|v''|+|w''|} - \frac{1}{3} \quad , }
\end{equation}
\rev{where $a = 2/3$ corresponds to having all turbulence kinetic energy in the streamwise velocity component $u''$, whereas $a = -1/3$ corresponds to having no energy in the streamwise $u''$ component. In figure~\ref{figure:2D_iso} (a) and (b) we show the $yz$-plane averaged anisotropy $\left< a \right>_{yz}$ as a function of the dimensionless mixing zone coordinate $\xi$ and time from Miranda and INCA. 
The dimensionless mixing zone coordinate $\xi$ is defined as }
\begin{equation}
 \rev{ \xi = \frac{x - x^*(t)}{\delta_x(t)} \quad , }
\end{equation}
\rev{with $x^*(t)$ being the x-location where $4(1-\phi(x,t))\phi(x,t)$ is maximal.} \par 
\rev{The light gas side of the mixing zone remains more anisotropic than the heavy gas side but with a homogeneous anisotropy distribution after re-shock on either side. The volume averaged anisotropy in the inner mixing zone $\left< a \right>_{xyz}$ is shown in figure~\ref{figure:2D_iso} (c). No full recovery of isotropy of the mixing zone is achieved, and the re-shock does not significantly contribute in the sense of the volume averaged quantity $\left< a \right>_{xyz}$, but leads to a stratified anisotropy distribution  around the centre of the mixing zone. After $t \approx 4.5~ms$ an asymptotic limit of $\left< a \right>_{xyz} \approx 0.04$ is reached, which temporally coincides with the onset of the self-similar decay of $\tke$, see figures~\ref{figure:TKE_loglog} and \ref{figure:2D_iso}. The positive value of $\left< a \right>_{xyz}$ implies that the streamwise component $u''$ remains, despite re-shock, the dominant velocity component throughout the simulation time. \citet{Lombardini2012} also found a temporal asymptotic limit of the isotropisation process in their simulations. \citet{Grinstein2011} observed that the velocity fluctuations in the mixing zone are more isotropic when the initial interface perturbations also include short wavelengths in which case the authors nearly recovered full isotropy. When \citet{Grinstein2011} used long wavelength perturbations the mixing zone remained anisotropic except for a narrow range on the heavy gas side.} \par
\rev{In order to quantify the homogeneity of mixing we calculate the density self-correlation \citep{Besnard1992}}
\begin{equation}
 \rev{ \left< b \right>_{yz}(\xi,t) = \left< - \left( \frac{1}{\rho}\right)'' \rho'' \right>_{yz} = \left< \frac{1}{\rho} \right>_{yz} \left< \rho\right>_{yz} - 1 \quad , }
\end{equation}
\rev{ which is non-negative. $\left< b \right>_{yz} = 0$ corresponds to homogeneously mixed fluids with constant pressure and temperature. Large values indicate spatial inhomogeneities in the respective yz-plane. The density self-correlation has gained some attention in recent years and was subject in several experimental investigations of the RMI, see \citet{Balakumar2012, Balasubramanian2012, Balasubramanian2013, Orlicz2013, Tomkins2013, Weber2014}.} \par
\begin{figure}
  \centering
    \includegraphics[width=1\textwidth]{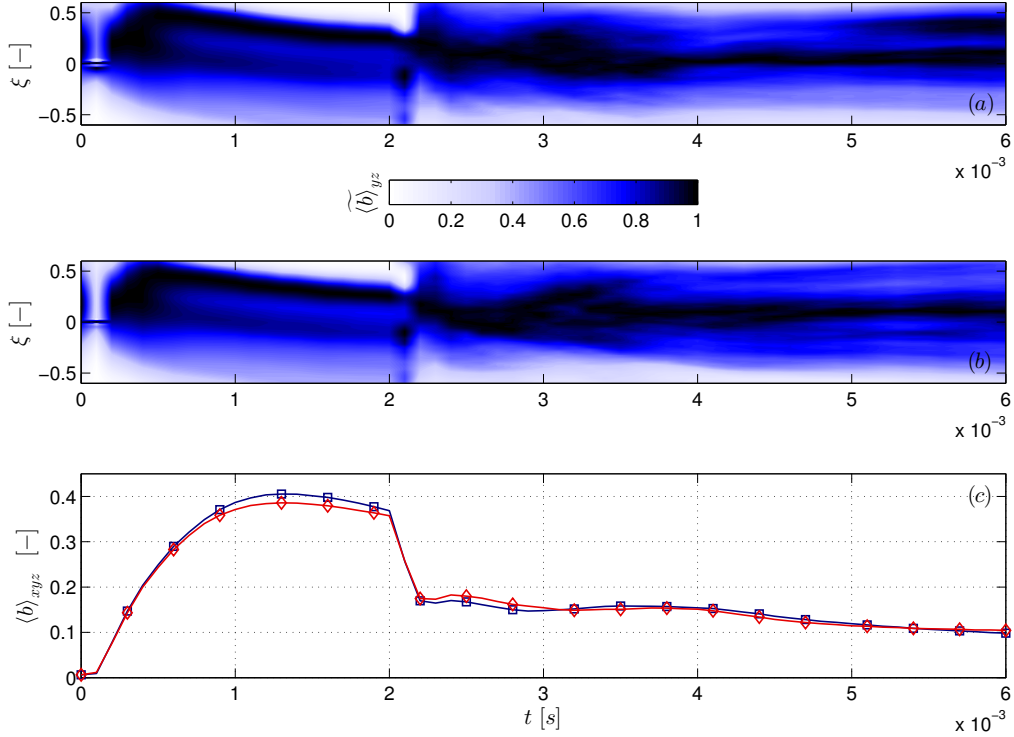}
  \caption{~Normalised density self-correlation $\widetilde{\left< b \right>}_{yz}$ as a function of the dimensionless mixing zone coordinate $\xi$ and time from Miranda (a) and INCA (b). The volume averaged density self-correlation $\left< b \right>_{xyz}$ of the inner mixing zone is shown in (c) from Miranda (squares) and INCA (diamonds). All data are taken from the finest grid ($512$).~}
  \label{figure:2D_b}
\end{figure}

\rev{Figure~\ref{figure:2D_b} (a) and (b) shows the density self-correlation normalised by the maximal value at time $t$ 
\begin{equation}
 \widetilde{\left< b \right>}_{yz} = \left< b \right>_{yz}(\xi,t) / \max(\left< b \right>_{yz})(t)
\end{equation}
as a function of the dimensionless mixing zone coordinate $\xi$ and time from Miranda and INCA. The largest values of $\widetilde{\left< b \right>}_{yz}$ are found around the centre of the mixing zone slightly shifted towards the heavy-gas side. $\widetilde{\left< b \right>}_{yz}$ peaks around the region where mixing between light and heavy gas occurs and tends to zero outside the mixing region, towards the respective pure gas side. \citet{Weber2014} observed in their experiment that the peak of the density self-correlation is initially shifted towards the light-gas side, but moves towards the center of the mixing zone with increasing time.} \par
\rev{In contrast to the anisotropy, where the re-shock does not contribute to the isotropisation and which levels out after $t \approx 4.5~ms$, the mixing zone becomes significantly more homogeneous after re-shock as can be observed from the temporal evolution of the volume average of the density self-correlation in the inner mixing zone $\left< b \right>_{xyz}$. Following re-shock the fluids become more and more mixed, see figure~\ref{figure:2D_b} (c) with a value of $\left< b\right>_{xyz} \approx 0.13$ at the latest time. The measured values of the density self-correlation in the single shock-interface interaction experiment of \citet{Weber2014} at $\Ma = 2.2$ are in good agreement with our simulated values at late times $\mathcal{O}(0.1)$, whereas at the lower Mach number $\Ma = 1.6$ \citet{Weber2014} observed a more inhomogeneous mixing zone $\mathcal{O}(0.2)$. These values are significantly larger than those measured for instance in the shock-gas-curtain experiments of \citet{Orlicz2013} and \citet{Tomkins2013}. }
\subsection{Spectral quantities}
From homogeneous isotropic turbulence it is well known that vorticity exhibits coherent worm-like structures with diameter on the order of the Kolmogorov length scale and of a length that scales with the integral scale of the flow. The work of \citet{Jimenez1992} suggests that theses structures are especially intense features of the background vorticity and independent of any particular forcing that generates the vorticity. In contrast to forced homogeneous isotropic turbulence where self-similar stationary statistics are achieved, shock-induced turbulent mixing is an inhomogeneous anisotropic unsteady decay phenomenon. Nevertheless, homogeneous isotropic turbulence is used as theoretical framework for most of the numerical analysis of RMI. However, it is unclear at what time and at what locations the mixing zone exhibits the appropriate features and if homogeneous isotropic turbulence is achieved at all. \rev{A fully isotropic mixing zone is never obtained, as the anisotropy, even though decreasing with time, reaches an asymptotic limit at $t \approx 4.5~ms$.} \par
\begin{figure}
  \centering
    \includegraphics[width=1\textwidth]{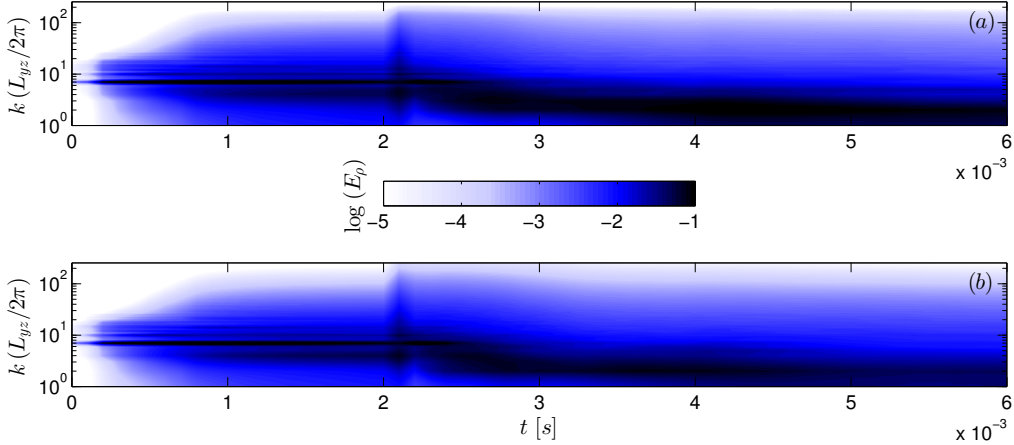}
  \caption{~Power spectra of density from Miranda (a) and INCA (b) as a function of wavenumber $k \left( L_{yz} / 2 \pi \right)$ and time. The data are taken from the finest grid ($512$).~}
  \label{figure:E_dens_kt}
\end{figure}

 
\rev{The temporal evolution of the initial perturbation is depicted in figure~\ref{figure:E_dens_kt}. Before re-shock the dominant modes of the initial perturbation slowly break down. After re-shock, however, the additional vorticity deposited during the second shock-interface interaction rapidly destroys structures generated by the initial perturbation and initial shock, leading to a self-similar decay after $t \approx 4~ms$.} \par
\citet{Thornber2010} and \citet{Thornber2012} found, formally in the limit of infinite Reynolds numbers, a persistent $k^{-3/2}$ scaling of the turbulence kinetic energy spectrum as well as a $k^{-3/2}$ spectrum with a $k^{-5/3}$ spectrum at high wavenumbers that covers more and more of the spectrum as time proceeds. Furthermore, the same authors \citep{Thornber2011} found (depending on the initial conditions) a $k^{-5/3}$ or a $k^{-2}$ scaling range after re-shock. Long after re-shock however, these scalings return to a $k^{-3/2}$ scaling at intermediate scales and to a $k^{-5/3}$ scaling at high wavenumbers, close to the cut-off wavenumber. The authors evaluated the radial spectra either in the centre of the mixing zone or averaged over a fixed number of yz-planes within the mixing zone.
A different scaling behaviour was observed by \citet{Lombardini2012} and \citet{Hill2006} who found in their multicomponent LES at finite Reynolds numbers a $k^{-5/3}$ scaling in the centre of the mixing zone. Whereas \citet{Cohen2002} found a $k^{-6/5}$ scaling range for the single-shock RMI averaged over four transverse slices within the mixing zone. In a recent experimental investigation of a shock accelerated shear-layer \citet{Weber2012} showed a $k^{-5/3}$ inertial range followed by an exponential decay in the dissipation range of the scalar spectrum. This result was numerically reproduced by \citet{Tritschler2013b}. Here, the authors averaged over a pre-defined $\imz$. \par
\begin{figure}
  \centering
    \includegraphics[width=1\textwidth]{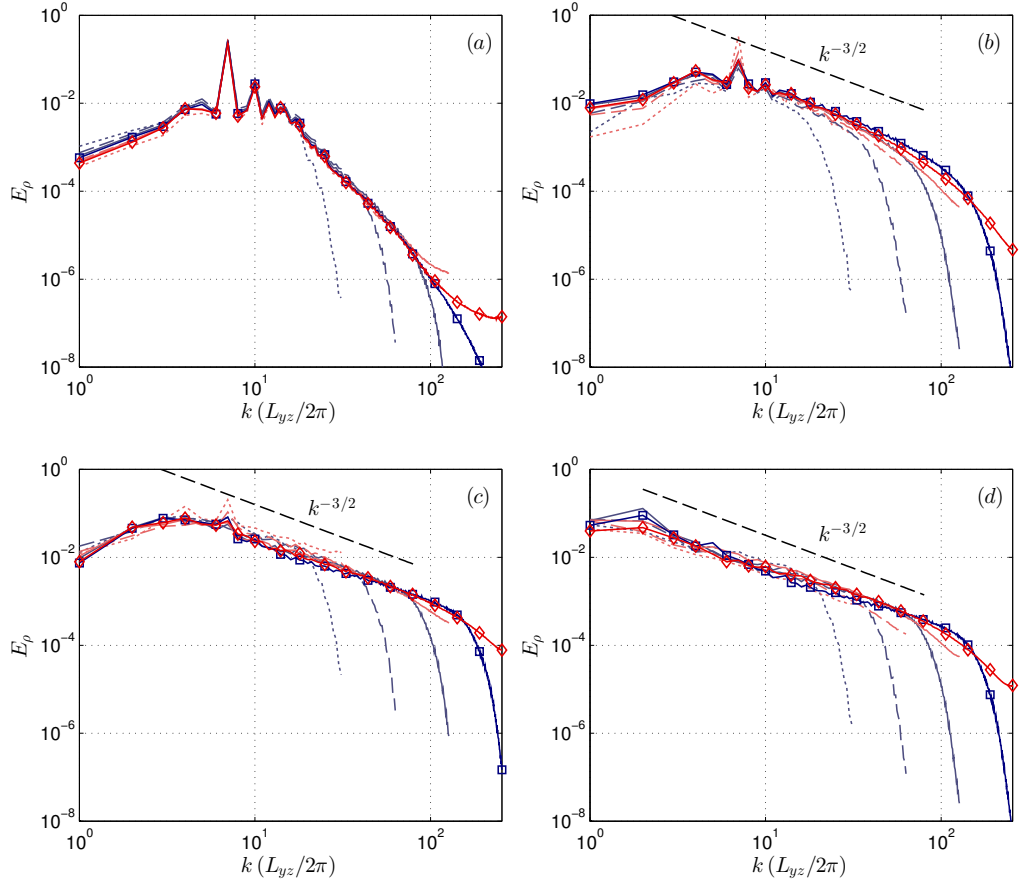}
  \caption{~Power spectra of density from Miranda (blue/dark grey) and INCA (red/light grey) before re-shock (a) $t=0.5~ms$ and (b) $t=2~ms$ and after re-shock (c) $t=2.5~ms$ and (d) $t=6~ms$. 
The different resolutions are represented as dotted line ($64$), dashed line ($128$), solid line ($256$) and solid line with open squares for Miranda and open diamonds for INCA ($512$).~}
  \label{figure:E_dens}
\end{figure}
\begin{figure}
  \centering
    \includegraphics[width=1\textwidth]{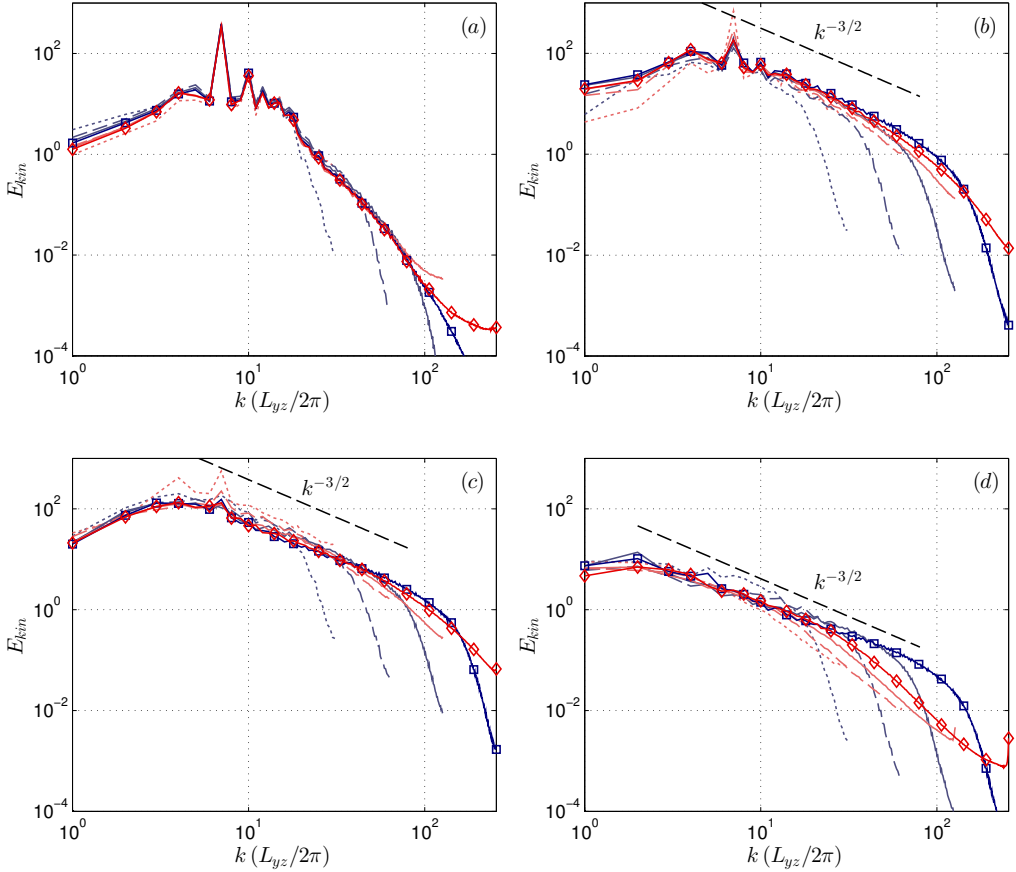}
  \caption{~Spectra of turbulence kinetic energy from Miranda (blue/dark grey) and INCA (red/light grey) before re-shock (a) $t=0.5~ms$ and (b) $t=2~ms$ and after re-shock (c) $t=2.5~ms$ and (d) $t=6~ms$. The different resolutions are represented as dotted line ($64$), dashed line ($128$), solid line ($256$) and solid line with open squares for Miranda and open diamonds for INCA ($512$).~}
  \label{figure:E_kin}
\end{figure}
All spectra shown in this section are radial spectra with a radial wavenumber that is defined as $k = \left( k_y^2 + k_z^2 \right)^{1/2}$. The radial spectra are averaged over all yz-planes within the $\imz$ in the x-direction that satisfy the condition in (\ref{eq:imz}). \s{It is worth noting that the chosen threshold defining the $\imz$ ($\ge 0.9$) affects the observed scaling and results in a slightly steeper inertial range than the $k^{-5/3}$ scaling. Increasing the threshold to unity the spectra recover the $k^{-5/3}$ scaling observed by Tritschler~\etal. Given the inhomogeneity and anisotropy of the mixing zone the radial spectra are clearly x-dependent.} \par
The radial power spectra of density are plotted in figure~\ref{figure:E_dens}, where figures~\ref{figure:E_dens}(a) and (b) show the spectra before and figures~\ref{figure:E_dens}(c) and (d) after re-shock. Power spectra of density and mass fraction concentration (not shown) show a close correlation, even though they are not directly related as the mass fractions are constrained to be between zero and one. \par
Before re-shock, the dominant initial modes slowly break down and redistribute energy to smaller scales. Re-shock causes additional baroclinic vorticity production with inverse sign that results in a  destruction process of the pre-shock structures\rev{, see also figure~\ref{figure:E_dens_kt}}. This process in conjunction with a vorticity deposition that is one order of magnitude larger than the pre-shock deposition leads to rapid formation of complex disordered structures, which eliminates most of the memory of the initial interface perturbation as can be seen in figures~\ref{figure:E_dens_kt}, \ref{figure:E_dens} and \ref{figure:E_kin}. \citet{Schilling2007} reported that during re-shock vorticity production is strongly enhanced along the interface where density gradients and misalignment of pressure and density gradients is largest. The vorticity deposited by the re-shock transforms bubbles into spikes and vice versa, which subsequently results in more complex and highly disordered structures. \par
\rev{At late times the power spectra of density appear to be more shallow than $k^{-3/2}$, and rather approach $k^{-6/5}$ as was found by \citet{Cohen2002}.}
\begin{figure}
  \centering
    \includegraphics[width=1\textwidth]{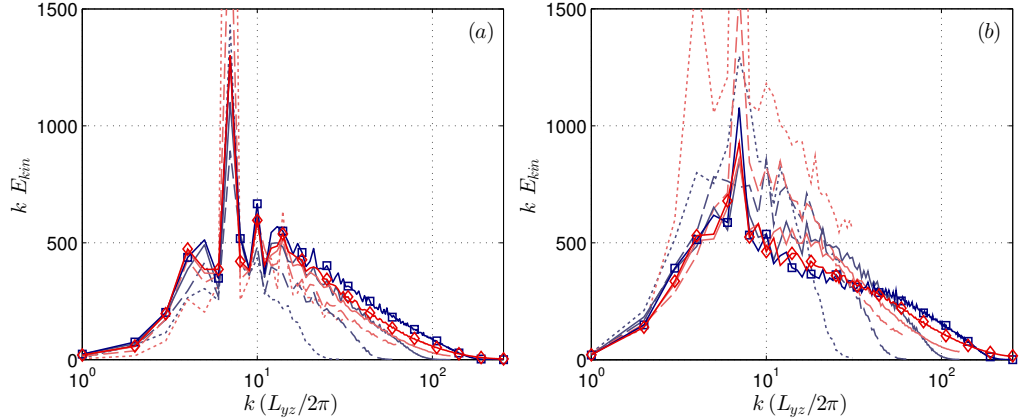}
  \caption{~Scaled spectra of turbulence kinetic energy from Miranda (blue/dark grey) and INCA (red/light grey) before re-shock (a) $t=2~ms$ and (b) $t=2.5~ms$. 
The different resolutions are represented as dotted line ($64$), dashed line ($128$), solid line ($256$) and solid line with open squares for Miranda and open diamonds for INCA ($512$).~}
  \label{figure:E_kin_comp}
\end{figure}
The smallest length scale in scalar turbulence is the Batchelor scale. For isotropic turbulence and Schmidt numbers of order unity it has the same order of magnitude as the Kolmogorov microscale $\lambda_B \approx \eta$. Therefore, the $\tke$ spectra are closely correlated with the scalar power spectra. Figure~\ref{figure:E_kin} shows the spectra of $\tke$ before and after re-shock.  The significant increase in $\tke$ is mainly due to the interaction of the enhanced small scale structures with comparatively steep density gradients and the reflected shock wave. The re-shock at $t \approx 2~ms$ leads to a self-similar lifting of the spectrum, see figure~\ref{figure:E_kin}. The destruction process of the vortical structures initiated by the re-shock leads to the formation of small scales, which rapidly remove the memory of the initial condition. The intense fluctuating velocity gradients past re-shock are rapidly smoothed out by viscous stresses. This results in a fast decay of the turbulence kinetic energy following the first $\approx 0.5~ms$ after re-shock, see figure~\ref{figure:TKE_Enstrophy} and figures~\ref{figure:E_kin}(c) and (d). \par
The sharp drop-off of the spectral energy in figures~\ref{figure:E_kin} and~\ref{figure:E_dens} in the Miranda data at high wavenumbers is due to the filtering operator of the numerical method. Opposite behaviour, that is an increase of spectral energy at the highest wavenumbers, is observed for the less dissipative INCA code, where the spurious behaviour at the non-resolved scales is mainly dispersive. \par
\begin{figure}
  \centering
    \includegraphics[width=1\textwidth]{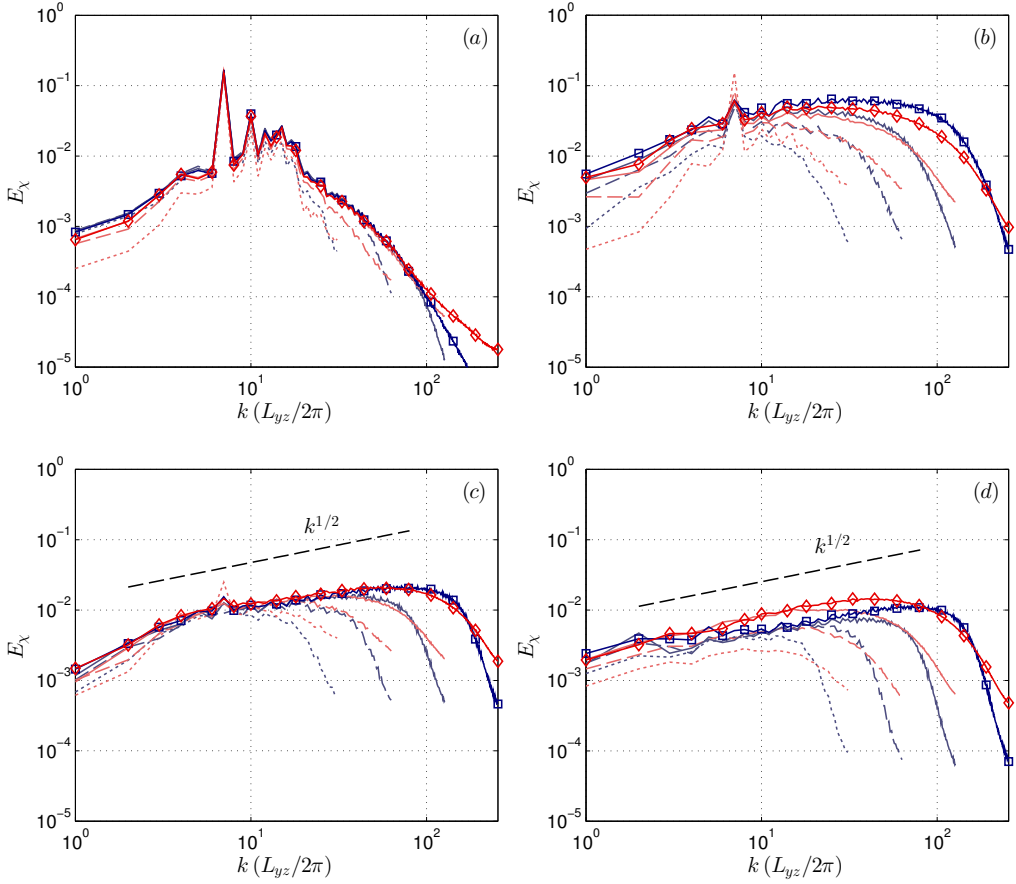}
  \caption{~Spectra of scalar dissipation rate from Miranda (blue/dark grey) and INCA (red/light grey) before re-shock (a) $t=0.5~ms$ and (b) $t=2~ms$ and after re-shock (c) $t=2.5~ms$ and (d) $t=6~ms$. The different resolutions are represented as dotted line ($64$), dashed line ($128$), solid line ($256$) and solid line with open squares for Miranda and open diamonds for INCA ($512$).~}
  \label{figure:E_MR}
\end{figure}
The \s{compensated}\rev{scaled} turbulence kinetic energy spectra $k E_{kin}(k)$ represent the effective energy contributed by each mode. Artifacts of the initial conditions still exist immediately before re-shock at $t=2.0~ms$ as can be seen in figure~\ref{figure:E_kin_comp}(a), where most energy is contained at mode $k (L_{yz}/2 \pi) = 7$.  At re-shock, baroclinic vorticity is deposited at the interface and the energy containing wavenumber range immediately \s{broadens}\rev{widens} as vortex stretching and tangling introduce new scales and higher vorticity.  This broader profile is plotted in figure~\ref{figure:E_kin_comp}(b) which clearly shows that the relative difference between the imposed initial length scale $k (L_{yz}/2 \pi) = 7$ and the remaining length scales (both larger and smaller) is vanishing. Indeed, as the mixing layer fully transitions to turbulence, \s{the flow becomes increasingly independent of the initial conditions}\rev{the flow reaches a self-similar state where the memory of initial perturbations is lost.} \par
\s{The spectra of the scalar dissipation rate $\chi$ in figure~\ref{figure:E_MR} quickly becomes broadband after the initial shock impact, see figure~\ref{figure:E_MR}(b), and develops an inertial subrange with a slope approximately between $k^{-1/2}$ and $k^{-1/4}$ which persists throughout the remainder of the simulation.}\rev{The spectra of the scalar dissipation rate $\chi$ in figure~\ref{figure:E_MR} quickly builds up in the cut-off wavenumber range after the initial shock impact, see figure~\ref{figure:E_MR}(b).} After re-shock and at late time, see figure~\ref{figure:E_MR}(c)-(d), the inertial subrange broadens to wavenumbers where numerical dissipation damps out structures. \rev{The inertial range is observed to scale with $k^{1/2}$ after re-shock, which is consistent with the $k^{-3/2}$ scaling observed for $E_\rho$ and $E_{kin}$.} For the resolved wavenumbers, there is good agreement between both codes at the finest two resolutions. \rev{Differences observed in figure~\ref{figure:E_MR} (b) are also reflected in figures~\ref{figure:MMF_TMR} and \ref{figure:2D_b}. A sharper material interface and the associated segregation of the fluids lead to a higher scalar dissipation rate ($\chi$), whereas at late times the difference in the scalar dissipation rate does not significantly influence the mixing measures $\left< b \right>_{xyz}$ and $\Theta$.} \par
\begin{figure}
  \centering
    \includegraphics[width=1\textwidth]{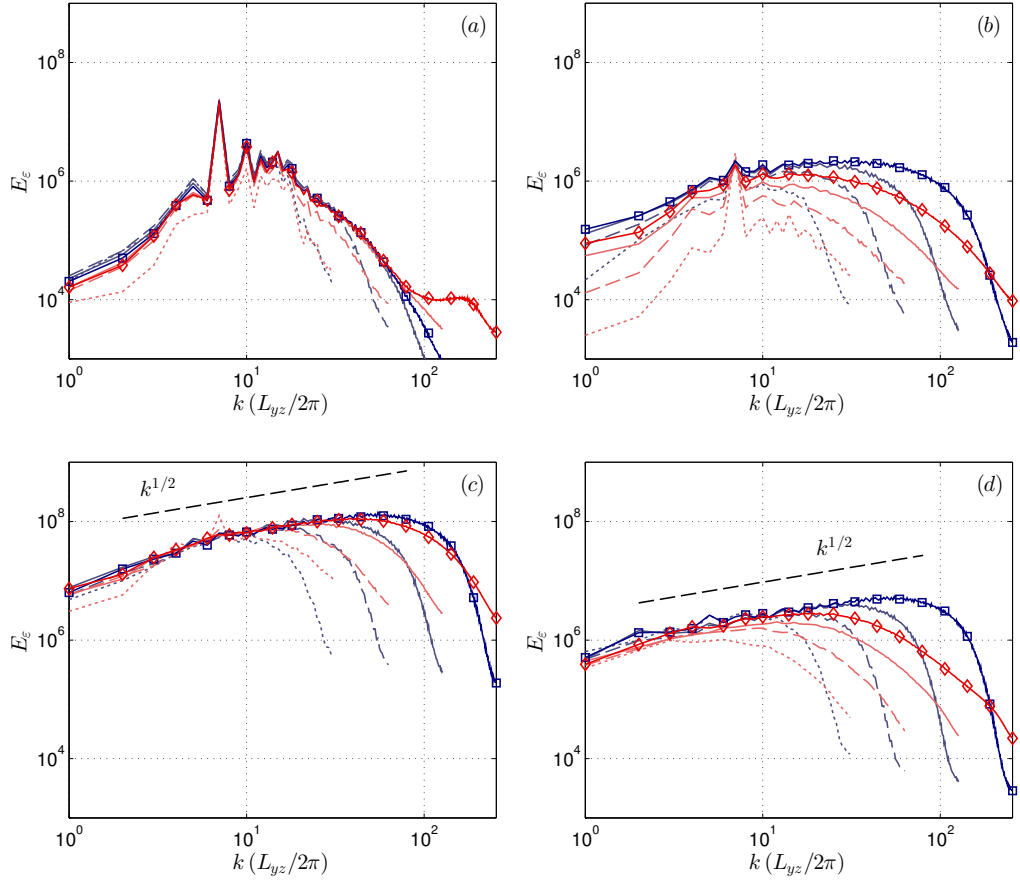}
  \caption{~Spectra of enstrophy from Miranda (blue/dark grey) and INCA (red/light grey) before re-shock (a) $t=0.5~ms$ and (b) $t=2~ms$ and after re-shock (c) $t=2.5~ms$ and (d) $t=6~ms$. 
The different resolutions are represented as dotted line ($64$), dashed line ($128$), solid line ($256$) and solid line with open squares for Miranda and open diamonds for INCA ($512$).~}
  \label{figure:E_enst}
\end{figure}
Larger quantitative differences are observed in the power spectra of enstrophy shown in figure~\ref{figure:E_enst}. Immediately after either of the shock-interface interactions the quantitative agreement between the predicted enstrophy levels is excellent, see figures~\ref{figure:E_enst}(a) and (c). \rev{The observed scalings of the inertial range following re-shock are predicted consistently and agree with the inertial range scalings for the scalar dissipation rate $k^{1/2}$.} However, the temporal decay of the \rev{small-scale} enstrophy is significantly different \s{in each of the codes}\rev{for either code} as can be seen immediately before re-shock and long after re-shock in figures~\ref{figure:E_enst}(b) and~\ref{figure:E_enst}(d), respectively. \par
\begin{figure}
  \centering
    \includegraphics[width=1\textwidth]{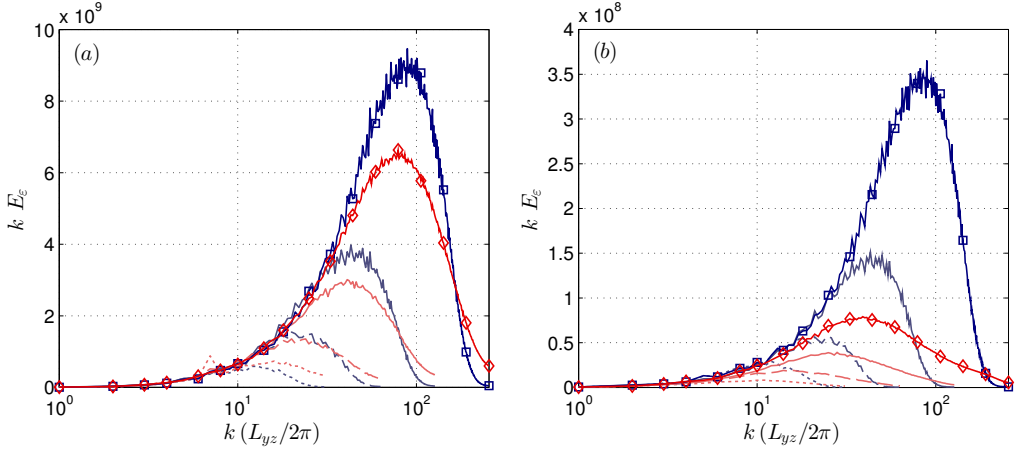}
  \caption{~Scaled spectra of enstrophy from Miranda (blue/dark grey) and INCA (red/light grey) after re-shock (a) $t=2.5~ms$ and (b) $t=6~ms$. The different resolutions are represented as dotted line ($64$), dashed line ($128$), solid line ($256$) and solid line with open squares for Miranda and open diamonds for INCA ($512$).~}
  \label{figure:E_enst_comp}
\end{figure}

In isotropic homogeneous turbulence, the \s{compensated}\rev{scaled} spectra of the enstrophy, see figure~\ref{figure:E_enst_comp}, has a single peak at the wavenumber where the dissipation range begins.  Therefore, under grid refinement this peak will shift to higher wavenumbers and magnitudes as smaller scales are captured. The peak at $k (L_{yz}/2 \pi) = 7$ is associated with the initial perturbation and disappears after re-shock as the flow becomes turbulent, see figure~\ref{figure:E_enst_comp}. Good agreement for lower wavenumbers is observed between codes and resolutions. Larger differences are observed at high wavenumbers where the dependence on numerical dissipation is greatest.  At $t=2.5~ms$ the peak in the \s{compensated}\rev{scaled} enstrophy spectra is at $k (L_{yz}/2\pi) \approx 85$ for both codes at the highest resolution.  Later, at $t=6.0~ms$ this peak has shifted to $k (L_{yz}/2\pi) \approx 40$ in INCA, whereas in Miranda, there is no apparent shift, although both have substantially decayed in magnitude. \par
As RMI is a pure decay process after re-shock, differences in the numerical approach become most apparent at late times. The numerical models of this study predict different turbulence decay rates as is evident from differences in the enstrophy spectrum, figure~\ref{figure:E_enst} and figure~\ref{figure:E_enst_comp}, and in $\tke$, figure~\ref{figure:TKE_loglog}. 
The differences in enstrophy ($\varepsilon$) and scalar dissipation rate ($\chi$) have a qualitative effect that becomes apparent in the fine scale structures of figure~\ref{figure:contour} at $t=6~ms$. Although INCA resolved less scales with smaller enstrophy levels, it does resolve steeper mass fraction gradients, which is reflected in the higher $\chi$ and higher levels of $E_{\chi}$ \s{over all wavenumbers}. Although it is unclear which dissipation rate (scalar or kinetic) \s{is most influential in}\rev{has most effect on} the mixing process, both \s{will have significant impact}\rev{are important} \citep{Dimotakis2000}. 
\subsection{Probability density functions}
The bin size for computing discrete probability density function (pdf) is defined as \s{$\Delta \varphi = [ \max(\varphi) - \min(\varphi)]/N_b$}\rev{$\Delta \varphi = [ \varphi_{max} - \varphi_{min}]/N_b$} for a quantity $\varphi(x,y,z,t)$. The number of bins for all quantities and all grid-resolutions is $N_b = 64$. Each discrete value of $\varphi$ is distributed into the bins, yielding a frequency $N_k$ for each bin. The $pdf$ is then defined by
\begin{equation}
 P_k(\varphi,t) = \frac{N_k}{\Delta \varphi N} \quad,
\end{equation}
such that $\sum_{k=1}^{N_k} P_k \Delta \varphi = 1$ with $N$ as the total number of cells in the $\imz$ \rev{that fall within the range $\varphi_{min} \le \varphi \le \varphi_{max}$}. The limits $\varphi_{max}$ and $\varphi_{min}$ are held constant for all resolutions and times. \par
\begin{figure}
  \centering
    \includegraphics[width=1\textwidth]{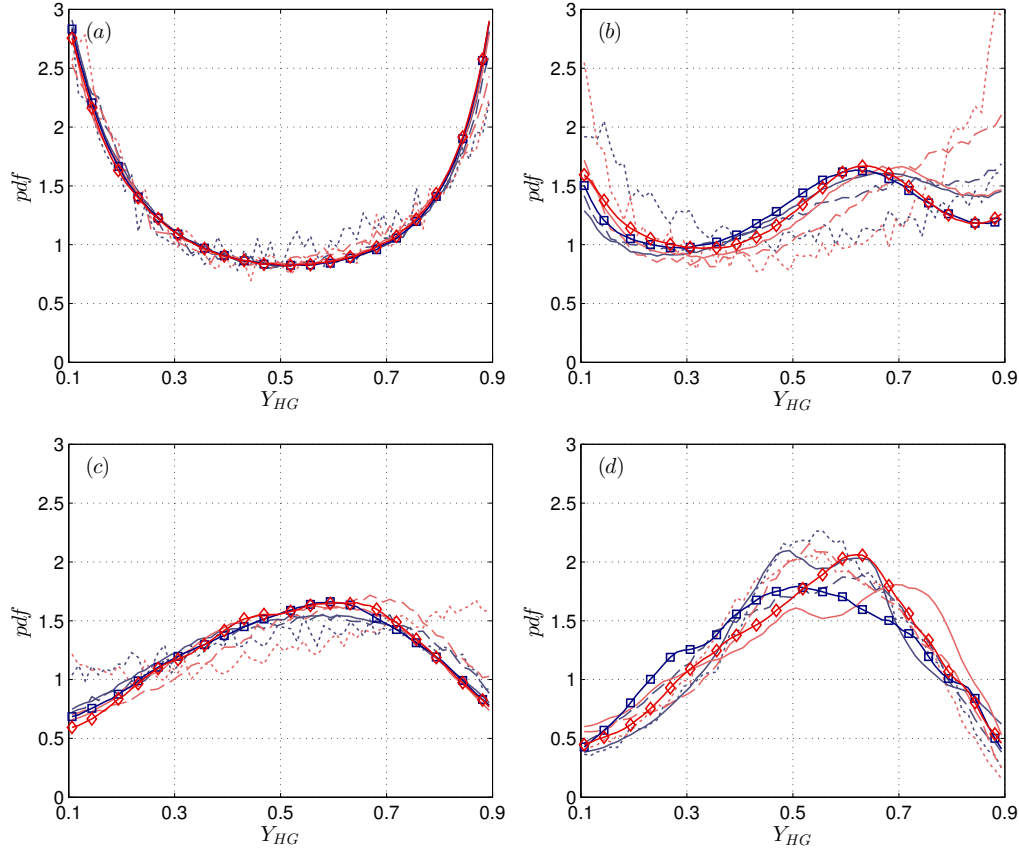}
  \caption{~Probability density function of $Y_{HG}$ from Miranda (blue/dark grey) and INCA (red/light grey)  before re-shock (a) $t=0.5~ms$ and (b) $t=2~ms$ and after re-shock (c) $t=2.5~ms$ 
and (d) $t=6~ms$. The different resolutions are represented as dotted line ($64$), dashed line ($128$), solid line ($256$) and solid line with open squares for Miranda and open diamonds for 
INCA ($512$).~}
  \label{figure:PDF_Y_SF6}
\end{figure}

The $pdf$ of the heavy-gas mass fraction is constrained to be $0.1\le Y_{HG} \le 0.9$. Figure~\ref{figure:PDF_Y_SF6} shows $pdf$ at times before re-shock ($t=0.5~ms$, $t=2~ms$) and 
following re-shock ($t=2.5~ms$, $t=6~ms$). From figure~\ref{figure:PDF_Y_SF6}(a) it is evident that at early times following the initial shock-interface interaction the $\imz$ consists mostly of segregated fluid as the large peaks at the $pdf$ bounds indicate. Before re-shock, interspecies mixing is largely dominated by the inviscid linear and nonlinear entrainment. Molecular diffusion processes have not yet had enough time to act, see figure~\ref{figure:PDF_Y_SF6}(b). Following re-shock, a fundamental change in the $pdf$ of $Y_{HG}$ ($P(Y_{HG})$) is observed, see figures~\ref{figure:PDF_Y_SF6}(c) and (d). The additional vorticity deposited by the re-shock leads to rapid formation of small and very intense vortical structures that lead to very effective mixing and destruction of the initial interface perturbation. The $pdf$ takes a uni-modal form at $t=2.5~ms$ as it was also reported by \citet{Hill2006}. The peak value, however, is not as well correlated with the average value of the mixture mass fraction as it was reported by \citet{Hill2006}. With our data the peak value is slightly shifted towards the heavy-gas side centered around $Y_{HG} \approx 0.6$. The degree of convergence between codes and resolutions is reassuring at $t=2.5~ms$. Note that $P(Y_{HG})$ is a very sensitive measure of the light-heavy gas mixing. \par
The rarefaction wave at $t \approx 3.2~ms$ does not significantly contribute to the mixing, as it is not as pronounced as found in comparable investigations \citep{Grinstein2011, Hill2006}. Long after re-shock the mixing process continues, which is reflected in narrower tails of $P(Y_{HG})$. The peak value of $Y_{HG}$ predicted by Miranda now coincides with the average value of the mixture mass fraction. In the INCA results this value remains slightly shifted towards the heavy gas side. \s{The bimodal character of $P(Y_{HG})$ reported by Hill~\etal, however, is not observed on the finest grid.}\rev{However, the bimodal character of $P(Y_{HG})$ reported by \citet{Hill2006}, who used $air$-$SF_6$ as light-heavy gas, is not observed on the finest grid.} Despite the strong mixing past re-shock the turbulent mixing zone remains inhomogeneous until the end of the simulation time, \rev{which makes the observed $pdf$ very sensitive to their location of evaluation within the mixing layer.} \par
The $pdf$ of the normalised vorticity is constrained between $0 \le \widetilde{\omega} \le 0.8$ with $\widetilde{\omega} = \omega \left( \lambda_L / v_s \right)$, where $v_s$ is the initial shock velocity and $\lambda_L$ is a characteristic length scale of the perturbations taken as $\lambda_L = L_{yz}/\widetilde{k}_{\max}$ where $L_{yz}$ is the width of the domain in the transverse direction and $\widetilde{k}_{\max} = k_{max} ( L_{yz}/2 \pi ) = 16$.\par
\begin{figure}
  \centering
    \includegraphics[width=1\textwidth]{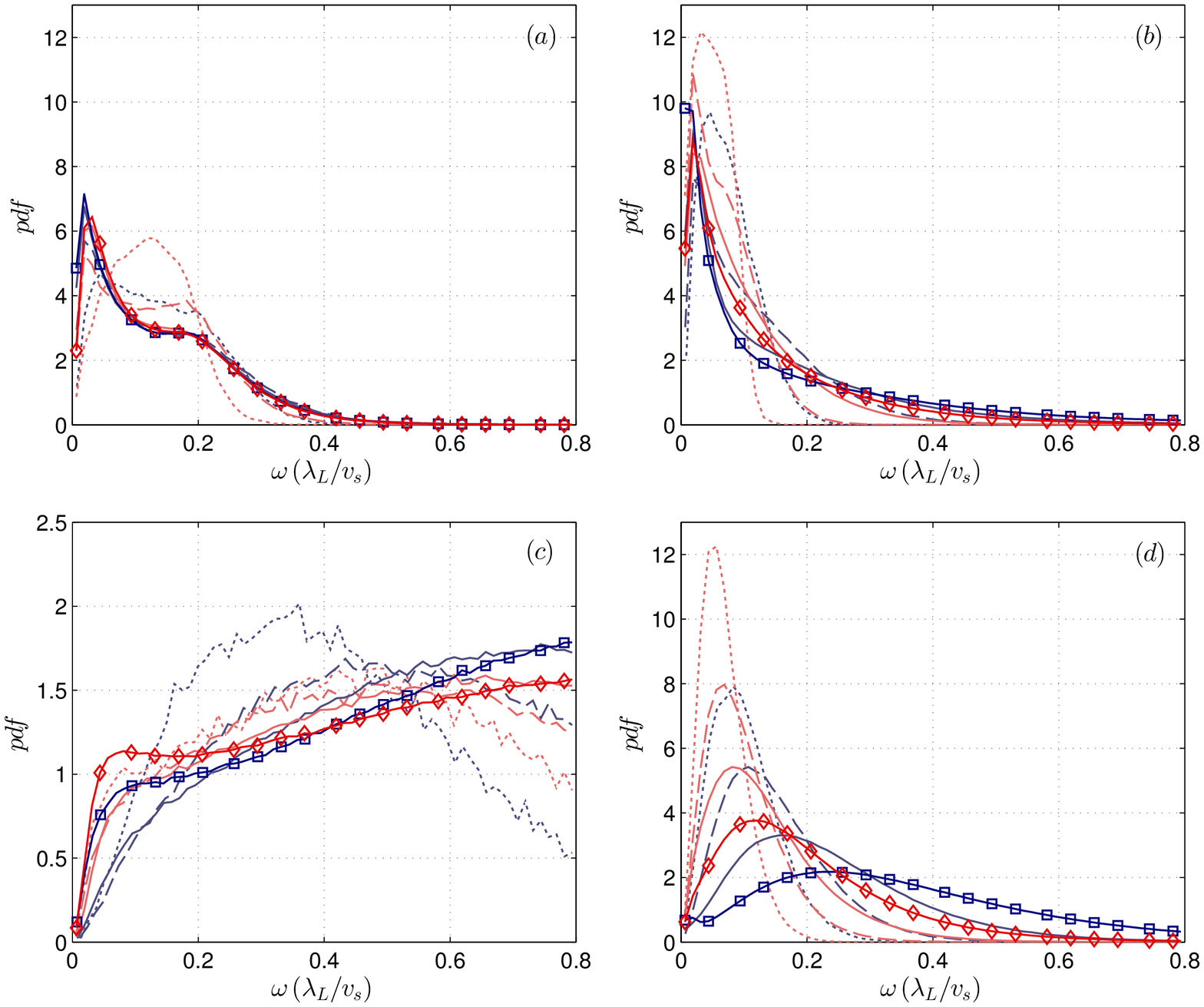}
  \caption{~Probability density function of $\omega \left( \lambda_L / v_s \right)$ from Miranda (blue/dark grey) and INCA (red/light grey)  before re-shock (a) $t=0.5~ms$ and (b) $t=2~ms$ and after re-shock (c) $t=2.5~ms$ and (d) $t=6~ms$. The different resolutions are represented as dotted line ($64$), dashed line ($128$), solid line ($256$) and solid line with open squares for Miranda and open diamonds for INCA ($512$).~}
  \label{figure:PDF_omega}
\end{figure}
Figure~\ref{figure:PDF_omega} shows the $pdf$ of the normalised vorticity $P(\tilde{\omega})$. Before re-shock, mixing is driven by weak large-scale vortices, see figures~\ref{figure:PDF_omega}(a) and (b). Following re-shock, however, structures with very intense vorticity develop with a dual-mode shape in $P(\tilde{\omega})$ at $t=2.5~ms$ on the finest grid. The early times after the second shock-interface interaction are again consistently predicted by both codes, figure~\ref{figure:PDF_omega}(c). Nevertheless, we observe larger differences for $P(\tilde{\omega})$ at $t=6~ms$. The peak and distribution of $P(\tilde{\omega})$ from Miranda are shifted towards larger values of $\tilde{\omega}$ as compared to INCA. This supports the previous observation that the vorticity decay is affected by the numerical approach\s{, especially when mixing is driven by intense small scale vortices}. \rev{The difference in the vorticity intensity observed in figure~\ref{figure:PDF_omega}, however, does not lead to noticeable differences for the integral mixing measures shown in figures~\ref{figure:MMF_TMR}(a) and \ref{figure:2D_b} or for the integral length scale in figure~\ref{figure:MLW_combined}. } 
\section{Conclusion}\label{sec:conclusion}
We have investigated the shock-induced turbulent mixing between a light ($N_2$, $O_2$) and heavy ($SF_6$, acetone) gas in highly resolved numerical simulations. The mixing was initiated by the interaction of a $\Ma = 1.5$ shock wave \s{that interacts} with a deterministic multimode interface. After the initial baroclinic vorticity deposition, the shock wave is reflected at the opposite adiabatic wall boundary. The reflected shock wave impacts the interface (re-shock) and deposits additional vorticity with \s{measured} enstrophy that is more than two orders of magnitude larger than that of the initial vorticity deposition. The \s{inversion}\rev{transformation} of spike structures into bubbles and vice versa in conjunction with a large increase in vorticity results in the formation of disordered structures which eliminate most of the memory of the initial interface perturbation.\par
\s{A goal of this study was to evaluate the proposed standardisation of the initial conditions for simulating the Richtmyer-Meshkov instability (RMI) by using two independent numerical approaches, Miranda and INCA, over a range of grid resolutions. The deterministic interface definition allows for spectrally identical initial conditions for different numerical models and grid resolutions. The direct comparison between the methods showed that larger energy containing scales are in excellent agreement. Moreover, most quantities showed a trend towards a grid-converged solution on the finest grid resolutions. The resolution study revealed insight at various stages of the RMI, as the flow transitions from large scale, non-linear entrainment, to fully developed turbulent mixing.} \rev{A proposed standardised initial condition for simulating the Richtmyer-Meshkov instability (RMI) has been assessed by two different numerical approaches, Miranda and INCA, over a range of grid resolutions. The deterministic interface definition allows for spectrally identical initial conditions for different numerical models and grid resolutions. A direct comparison shows that larger energy containing scales are in excellent agreement. Different subgrid-scale regularisation affect marginally resolved flow scales, but allow for a clear identification of a resolved scale-range that is unaffected by the subgrid-scale regularisation.}\par 
Mixing widths are nearly identical between the two approaches at the highest resolution. At lower resolutions, the solutions differ, and we found a minimum resolution of $\sim 400~\mu m$ to be necessary in order to produce reasonable late-time results. The initial mixing zone growth rate scaled with $\delta_x \sim t^{7/12}$, whereas long after re-shock the predicted growth rate was $\sim t^{2/7}$. The decay of turbulence kinetic energy was also found to be consistent and in good agreement between the approaches. The decay scaled with $\tke \sim t^{-10/7}$ at late times, that corresponds to a growth rate scaling of the energy-containing eddies of $\sim t^{2/7}$. The agreement in the large scales of the solution between the two approaches is striking and has not been observed before. \s{Therefore, the proposed standardisation is useful for the generation of high fidelity data sets to be used for physics exploration and model development.} \par
\s{The other purpose of this study was to elucidate differences in the behaviour of RMI and turbulent mixing that arises from the numerical method used in the LES methodology. A previous work on three-dimensional LES of RMI has examined numerical dependence only indirectly, c.f. Thornber~\etal~(2010), where a single shock RMI was studied between two codes. However, the initialisation was different between the codes and as the purpose of that study was not directly to measure numerical method dependence, comparison between codes and resolutions for mean, spectral and gradient based quantities was limited. An unprecedented level of detail, in terms of quantities examined across resolutions and numerical methods, has been achieved.}\rev{Previous work on three-dimensional LES of RMI has examined numerical dependence only indirectly. E.g.~\citet{Thornber2010} performed a code comparison of single-shock RMI. However, the initialisation was different for the two codes and the purpose was not to quantify the effect of different numerical methods. Accordingly, the comparison of results at different resolutions for mean, spectral and gradient based quantities was limited. With the current work we have presented for the first time a comprehensive quantitative analysis of numerical effects on RMI.} \par 
Results conclusively show that the large scales are in excellent agreement \s{between codes}\rev{for the two methods}. Differences are observed in the representation of the material interface. \s{This difference is reflected in the $SF_6$ mass fraction contour plots and in the molecular mixing fraction ($\Theta$).} We conclude that the numerical challenge, prior to re-shock, is to predict the large-scale non-linear entrainment and the associated interface sharpening. Under shear and strain the interface steepens and eventually becomes under-resolved with a thickness defined by the resolution limit of the numerical scheme. Therefore, the saturation of the interface thickness by the numerical method occurs later as the grid is refined. The molecular mixing fraction reached an asymptotic limit as $\Theta \approx 0.85$ after re-shock, which was already correctly calculated on grid-resolutions of $\sim 400~\mu m$. \rev{The $yz$-plane averaged anisotropy $\left< a\right>_{yz}$ revealed that the mixing zone exhibits a stratified anisotropy distribution with lower anisotropy on the heavy-gas side and higher anisotropy on the light-gas side. Moreover, the volume-averaged anisotropy $\left< a\right>_{xyz}$ approached an asymptotic limit of $\left< a\right>_{xyz} \approx 0.04$, implying that the fluctuating velocity component $u''$ remains the dominant component even after re-shock and that no full recovery of isotropy of the mixing zone is obtained. The density-self correlation has been investigated in order to better understand the mixing inhomogeneity in the mixing zone. The volume-averaged density self-correlation $\left< b\right>_{xyz}$ showed that the re-shock significantly increases mixing homogeneity approaching a value of $\left< b\right>_{xyz} \approx 0.13$ at the latest time.} \par
The spectra demonstrate a broad range of resolved scales, which are in very good agreement. \s{However, data have also shown that differences do exist in the fine scales and in quantities which are more affected by small scales.}\rev{Data also show that differences exist in the small-scale range.} The frequency dependence of the velocity and density fluctuations shows the existence of an inertial subrange and that the two approaches agree at lower frequencies. The observed spectral scalings were consistent among the methods \s{but had slightly steeper slopes}\rev{with $k^{-3/2}$.} \s{than the classical Kolmogorov scaling.} \s{This difference was traced back to the averaging procedure and the anisotropy of the mixing zone.} \par
\s{Metrics}\rev{Quantities} that are gradient dependent and therefore \rev{more} sensitive to small scales, such as the scalar dissipation rate and enstrophy, exhibit stronger dependence on  numerical method and grid resolution. \s{For the grid resolutions considered we did not observe grid convergence for these quantities.} The flow field shows visual differences for the fine-scale structures at late times. \rev{The $10^{th}$-order compact scheme and the explicit filtering and artificial fluid properties used in Miranda resolved more small scales in turbulence kinetic energy and enstrophy, whereas the $6^{th}$-order WENO based scheme used in INCA resolves more of the small-scale scalar flow features as observed in the spectra of density and scalar dissipation rate.  This result is somewhat intuitive given the numerics of the two codes. High-order compact methods are capable of resolving higher modes than explicit finite difference methods, \citep{lele:1992}. Given that the artificial shear viscosity in Miranda has only a small effect on the solution compared to the effect of the $8^{th}$-order filter, the primary difference, we conclude, of the resolving power between methods is due to the difference in order of accuracy and modified wavenumber profiles between the schemes. The compact finite difference method with high-order filtering appears to capture a broader range of dynamic scales at late times. } \par
The $pdf$ statistics of heavy gas mass fraction $Y_{HG}$ revealed that the $\imz$ remains inhomogeneous until the end of the simulation and that the peak probability is centered around $Y_{HG} \approx 0.6$ and thus is slightly shifted towards the heavy gas side. Although the overall quantitative agreement was very good, the $pdf$ of the vorticity  showed larger differences once intense small-scale vortical structures exist. The decay of vorticity differs accordingly between the numerical \s{models}\rev{methods}. \\

The authors gratefully acknowledge the Gauss Centre for Supercomputing e.V. (www.gauss-centre.eu) for providing computing time on the GCS Supercomputer SuperMUC at Leibniz Supercomputing Centre (LRZ, www.lrz.de). This work was performed under the auspices of the U.S. Department of Energy by Lawrence Livermore National Laboratory under contract number DE-AC52-07NA27344. VKT gratefully acknowledges the support of the TUM Graduate School. BJO thanks A. Cook and W. Cabot for valuable insight and for use of the Miranda code. 
\clearpage
\appendix
\section{Multicomponent mixing rules} \label{ap:multicomponent mixing rules}
The specific heat capacity of species $i$ is found by
\begin{equation}
 c_{p,i} = \frac{\gamma_i}{\gamma_i - 1} R_i \quad \mbox{,with} \quad R_i = \frac{R_{univ}}{M_i} \quad ,
\end{equation}
where $\gamma_i$ is the ratio of specific heats. The ratio of specific heats of the mixture follows as 
\begin{equation} \label{eq:mixture_gamma}
 \overline{\gamma} = \frac{\overline{c_p}}{\overline{c_p} - \overline{R}} \quad \mbox{, with} \quad  \overline{c_p} = \sum_i^N Y_i c_{p,i} \quad,
\end{equation}
where $Y_i$ is the mass fraction of species $i$ and $\overline{R}$ is the specific gas constant of the mixture with 
$\overline{R} = R_{univ}/\overline{M}$. The molar mass of the mixture is given by 
\begin{equation}
 \overline{M} = \left(\sum_i^N \frac{Y_i}{M_i} \right)^{-1} \quad .
\end{equation}
For the gas mixture Dalton's law $p= \sum_i p_i$ shall be valid with $p_i = \rho R_i T$.
The mixture viscosity $\overline{\mu}$ and the mixture thermal conductivity $\overline{\kappa}$ is calculated from \citep{Reid1987}
\begin{equation}
 \overline{\mu} = \frac{\sum_{i=1}^N \mu_i Y_i/M_i^{1/2}}{\sum_{i=1}^N Y_i/M_i^{1/2}} \qquad , \quad \overline{\kappa} = \frac{\sum_{i=1}^N \kappa_i Y_i/M_i^{1/2}}{\sum_{i=1}^N Y_i/M_i^{1/2}} \quad .
\end{equation}
The effective binary diffusion coefficients (diffusion of species $i$ into all other species) are approximated as \citep{Ramshaw1990}
\begin{equation} \label{eq:effective binary diffusion coefficient}
 D_i = \left( 1 - X_i \right) \left( \sum_{i \neq j}^N \frac{X_j}{D_{ij}}\right)^{-1} \quad ,
\end{equation}
where $X_i$ is the mole fraction of species $i$. (\ref{eq:effective binary diffusion coefficient}) ensures that the inter-species diffusion fluxes balance to zero. 
\section{Molecular mixing rules} \label{ap:molecular mixing rules}
The viscosity of a pure gas is calculated from the Chapman-Enskog model \citep{Chapman1990}
\begin{equation}
 \mu_i = 2.6693 \cdot 10^{-6} \frac{\sqrt{M_i T}}{\Omega_{\mu,i} \sigma_i^2} \quad ,
\end{equation}
where $\sigma_i$ is the collision diameter and $\Omega_{\mu,i}$ is the collision integral
\begin{equation}
 \Omega_{\mu,i} = A(T_i^*)^B + C \exp{\{D T_i^*\}} + E \exp{\{F T_i^*\}} \quad,
\end{equation}
with $A =  1.16145$, $B = -0.14874$, $C =  0.52487$, $D = -0.7732$, $E =  2.16178$ and $F = -2.43787$ and $T_i^* = T / (\epsilon / k)_i$. 
$(\epsilon / k)_i$ is the Lennard-Jones energy parameter, where $\epsilon$ is the minimum of the Lennard-Jones potential and $k$ is the Boltzmann constant.\par
The thermal conductivity of species $i$ is defined by
\begin{equation}
 \kappa_i = c_{p,i} \frac{\mu_i}{Pr_i} 
\end{equation}
with $Pr_i$ as the species specific Prandtl number.\par
The mass diffusion coefficient of a binary mixture is calculated from the empirical law \citep{Reid1987}
\begin{equation}
 D_{ij} = \frac{0.0266}{\Omega_{D,ij}} \frac{T^{3/2}}{p \sqrt{M_{ij}} \sigma_{ij}^2}
\end{equation}
with the collision integral for diffusion
\begin{equation}
 \Omega_{D,ij} = A(T_{ij}^*)^B + C \exp{\{D T_{ij}^*\}} + E \exp{\{F T_{ij}^*\}} + G\exp{\{H T_{ij}^*\}} \quad,
\end{equation}
where $T_{ij}^* = T / T_{\epsilon_{ij}}$ and $A =  1.06036$, $B = -0.1561$, $C =  0.19300$, $D = -0.47635$, $E =  1.03587$, $F = -1.52996$, $G =  1.76474$, $H = -3.89411$ and
\begin{subeqnarray}
 M_{ij}             & = & \frac{2}{\frac{1}{M_i} + \frac{1}{M_j}} \\
 \sigma_{ij}        & = & \frac{\sigma_i + \sigma_j}{2} \\
 T_{\epsilon_{ij}}  & = & \sqrt{ \left( \frac{\epsilon}{k} \right)_i \left( \frac{\epsilon}{k} \right)_j  } \quad .
\end{subeqnarray}
The molecular properties of all species in the present study are given in Table~\ref{table:molecularproperties}.
\begin{table}
  \centering
    \begin{tabular}{ ccccc} 
    \hline
    Property                        & Nitrogen  & Oxygen    & $SF_6$     & Acetone\\ 
    $\left(\epsilon/k\right)_i~[K]$ & $82.0$    & $102.6$   & $212.0$    & $458.0$   \\ 
    $\sigma_i~[ \mathring{A}]$      & $3.738$   & $3.48$    & $5.199$    & $4.599$   \\ 
    $M_i~[g~mol^{-1}]$              & $28.0140$ & $31.9990$ & $146.0570$ & $58.0805$ \\ 
    $\gamma_i$                      & $1.4$     & $1.4$     & $1.1$      & $1.1$     \\ 
    $Pr_i$                          & $0.72 $   & $0.72 $   & $0.8$      & $0.8$     \\
    \end{tabular}
    \caption{~~Molecular properties of nitrogen, oxygen, $SF_6$ and acetone.~~}
  \label{table:molecularproperties}
\end{table}


\bibliographystyle{jfm}
\bibliography{references_040313}

\end{document}